\documentclass[a4paper,12pt]{article}

\usepackage{epsfig}
\usepackage{a4}
\usepackage{latexsym, amssymb} 
\usepackage{amsmath}
\newcommand{\lapprox}{%
\mathrel{%
\setbox0=\hbox{$<$}
\raise0.6ex\copy0\kern-\wd0
\lower0.65ex\hbox{$\sim$}
}}
\newcommand{\gapprox}{%
\mathrel{%
\setbox0=\hbox{$>$}
\raise0.6ex\copy0\kern-\wd0
\lower0.65ex\hbox{$\sim$}
}}

\newcommand{\ba}{\begin{array}}
\newcommand{\ea}{\end{array}}
\newcommand{\bd}{\begin{displaymath}}
\newcommand{\ed}{\end{displaymath}}
\newcommand{\be}{\begin{equation}}
\newcommand{\ee}{\end{equation}}
\newcommand{\bea}{\begin{eqnarray}}
\newcommand{\eea}{\end{eqnarray}}

\def\q2 {q^2}

\def\fb{\, {\rm fb}}

\def\met{E_T \hspace*{-1.1em}/\hspace*{0.5em}}

\def\gev{\, \, {\rm GeV}}
\def\neu{\tilde{\chi}_1^0}
\def\st{\sin\theta_{\tilde{\tau}}}
\def\ct{\cos\theta_{\tilde{\tau}}}
\def\th{\theta_{\tilde{\tau}}}
\def\mt{m_\tau}
\def\tla{\tilde{\tau}_1}
\def\tlb{\tilde{\tau}_2}
\def\tli{\tilde{\tau}_i}
\def\mla{m_{\tla}}
\def\mlb{m_{\tlb}}
\def\mli{m_{\tilde{\tau}_i}}

\def\YR{Y_R}
\def\YL{Y_L}
\def\B{\tilde{B}}
\def\mB{m_{\tilde{B}}}

\def\bt{\begin{table}}
\def\et{\end{table}}

\catcode`@=11 
\def \gsim{\mathrel{\mathpalette\@versim>}}
\def \lsim{\mathrel{\mathpalette\@versim<}}
\def \@versim#1#2{\lower0.4ex\vbox{\baselineskip\z@skip\lineskip\z@skip
     \lineskiplimit\z@\ialign{$\m@th#1\hfil##\hfil$%
     \crcr#2\crcr\sim\crcr}}}
\catcode`@=12 

\begin{document}

\begin{flushright}
{\small 
IPMU13-0167\\
KEK-TH-1661}
\end{flushright}

\begin{center}

{\large\bf 10 GeV neutralino dark matter and light stau in the MSSM}\\[15mm]
Kaoru Hagiwara\footnote{E-mail: kaoru.hagiwara@kek.jp},
Satyanarayan Mukhopadhyay \footnote{E-mail: satya.mukho@ipmu.jp}
and Junya Nakamura\footnote{E-mail: junnaka@post.kek.jp}\\ \bigskip
{\em $^{1,3}$KEK Theory Center and Sokendai, \\
Tsukuba, Ibaraki 305-0801, Japan\\ \bigskip
$^2$Kavli IPMU (WPI), The University of Tokyo,\\
Kashiwa, Chiba 277-8583, Japan.}
\\[20mm] 
\end{center}

\begin{abstract} 
It has recently been pointed out that a component of the observed gamma ray emission in the low-latitudes of Fermi Bubble has a spectral shape that can be explained by a 10 GeV dark matter (DM) annihilating to tau leptons with a cross-section of $2 \times 10^{-27} {~\rm cm^3}$/s. Motivated by this possibility, we revisit the annihilation of a 10 GeV neutralino DM in the MSSM via stau exchange. The required stau masses and mixing, consistent with LEP direct search and electroweak precision constraints, are correlated with a possible enhancement of the Higgs decay rate to two photons. We also explore the implications of such a scenario for DM relic density and the muon anomalous magnetic moment, taking into account the recent ATLAS bounds on the chargino and the first two generation slepton masses, as well as the constraints on the Higgsino fraction of a 10 GeV neutralino. 
\end{abstract}

\vskip 1 true cm

\newpage
\setcounter{footnote}{0}

\def\baselinestretch{1.5}
\section{Introduction}
The observation of two large gamma-ray lobes in the Fermi-LAT data, extending about $50^\circ$ above and below the Galactic Centre~\cite{Su}, has generated a lot of interest regarding its origin in possible astrophysical processes. Recently, by studying the variation of the gamma-ray spectrum in these Fermi Bubbles with Galactic latitude, it was argued in Ref.~\cite{HS-1} that far away from the Galactic plane, above latitudes of about $30^\circ$, the spectrum can be explained by inverse Compton scattering of the interstellar radiation with cosmic ray electrons. On the otherhand, within around $20^\circ$ of the Galactic plane, the spectrum is found to have a peak at energies of 1-4 GeV, which is a feature difficult to explain by the same mechanism with a realistic electron population~\cite{HS-1}. As possible explanations of the non-inverse Compton component, cosmic ray protons scattering with gas, a large population of millisecond pulsars and dark matter annihilations have been studied by comparing their predictions with the  low latitude emission,  after subtracting the inverse Compton component~\cite{HS-1,HS-2,Gordon,W-wimp,Huang}. The resulting gamma ray spectrum from dark matter (DM) annihilation is found to provide a good fit to the background subtracted low latitude data, using a generalized Navarro-Frenk-White profile for  DM distribution, with a slightly steeper inner slope of 1.2, and a local DM density of $0.4$ GeV/cm$^3$. The possible DM mass and annihilation rate include a 10 GeV DM annihilating to $\tau^+\tau^-$, with an annihilation rate $\langle \sigma v \rangle\sim 2 \times 10^{-27} {\rm cm}^3/{\rm s}$, or a 50 GeV DM annihilating to $b \bar{b}$, with  $\langle \sigma v \rangle \sim 8 \times 10^{-27} {\rm cm}^3/{\rm s}$~\cite{HS-1}. For details on the extraction of this possible signal, we refer the reader to Ref.~\cite{HS-1}, and for alternative possibilities of DM annihilation modes to Ref.~\cite{Gordon}.  The annihilating DM model discussed above can also account for gamma ray emissions with similar spectral features previously identified close to the Galactic Centre~\cite{HS-1,Hooper-10GeV}.

In the R-parity conserving minimal supersymmetric standard model (MSSM), the lightest neutralino, $\neu$, is a viable DM candidate, whose mass can range from a few GeV to several TeV. The neutralino DM in all mass range have been studied as a possible thermal relic in specific models of supersymmetry (SUSY) breaking as well as in a general phenomenological MSSM~\cite{Jungman,Bertone}. In particular, lower bounds on the mass of a light neutralino DM ($\neu$) have been obtained in the MSSM under various assumptions, taking into account constraints from accelerator, flavour physics and direct detection experiments. In general, although a light $\mathcal{O}(10 {~\rm GeV})$ neutralino is allowed by most constraints, the requirement of a sufficiently large thermally averaged annihilation cross-section at the epoch of its freeze-out in the early universe (so that DM is not over-abundant in the present epoch), brings in a certain amount of tension in  the MSSM parameter space. However, possible alternative thermal histories of the universe, for instance, the effect of late-time decay of a heavy particle like moduli into light states (without disturbing the successful big-bang nucleosynthesis), have also been discussed in various contexts, which can revive parameter points in which the DM relic density is otherwise predicted to be larger~\cite{Hooper-moduli}.

Motivated by the analysis of the gamma-ray data in the Fermi Bubble discussed above, and the possibility that light DM annihilation can provide a second emission mechanism necessary to explain the features of the non-inverse-Compton component, we revisit the case for a 10 GeV neutralino DM ($\neu$) annihilating to $\tau^+\tau^-$ in the context of the MSSM. The dominant contribution to this annihilation process in the MSSM comes from the exchange of scalar partners of tau leptons (stau), and therefore,  the requirement of an annihilation rate of $\langle \sigma v\rangle_0\sim 2 \times 10^{-27} {\rm cm}^3/{\rm s}$ constrains  the model parameters in the stau sector, namely the stau masses $\mla$, $\mlb$, and the mixing angle $\th$. Interestingly, the same set of MSSM parameters in the stau sector determine the dominant SUSY contribution to the Higgs boson ($h$) decay rate to two photons measured at the Large Hadron Collider (LHC). Both the $\neu$ annihilation rate to $\tau^+ \tau^-$ and the $h \rightarrow \gamma \gamma$ decay width are enhanced in presence of a light $\tla$, a large mixing  $\th$ between the stau current eigenstates and a large mass splitting $\Delta m_{\tilde{\tau}} = \mlb - \mla$. The central value of the $h \rightarrow \gamma \gamma$ signal strength measured by ATLAS is $1.6$ (although consistent at $2\sigma$ with the SM prediction and the CMS measurement),  and hence there is a possibility that evidence for new physics might show up in this channel in the future LHC runs. The primary goal of this paper is therefore to correlate these two observables and find out the parameter region in the stau sector consistent with both of them. 

The stau masses and mixing angle are, however, constrained by LEP direct search for stau pair production, as well as electroweak precision measurements. We revisit both these constraints in our study. Using  the LEP stau search limits, we clarify the mixing angle dependence of the bound on $\tla$ mass, and from the electroweak precision constraints, we determine the allowed splitting between the mass eigenstates $\tla$ and $\tlb$.  Although from LEP constraints on chargino mass and the $Z$-boson total invisible width, a $10$ GeV $\neu$ has to be dominantly bino-like, it can still have a small Higgsino fraction. Moreover, recent bounds on the chargino mass from the 8 TeV LHC, as well as the current upper bound on the Higgs boson invisible branching ratio obtained from a global fit of the LHC Higgs measurements can lead to more stringent constraints on  the Higgsino fraction of a 10 GeV neutralino, depending upon the values of some other SUSY parameters. After determining the allowed Higgsino fraction taking all these constraints into account, we comment on its implications for the $\neu$ annihilation cross-section to tau leptons. Combined with the current limits on the first two generation slepton masses from the ATLAS collaboration, this bound  on the Higgsino fraction also has implications for the relic density of the $\neu$, its direct detection rate, and the SUSY contribution to the anomalous magnetic moment of the muon $g-2$. 

This paper is organised as follows. In section~\ref{sec:constraints} we discuss the constraints on the relevant MSSM parameters. We study $\neu$ annihilation to $\tau^+\tau^-$ in the present universe and determine the stau masses and mixing necessary to reproduce the annihilation rate $\sigma v_0$ in section~\ref{sec:annihilation}. In section~\ref{sec:higgs}, we correlate these parameters with the $h \rightarrow \gamma \gamma$ decay rate, after including the stau loop contributions. Taking into account the current constraints on slepton masses, the relic density of a $10$ GeV $\neu$ is briefly discussed in section~\ref{sec:relic}, while the MSSM contribution to the muon $g-2$ in this context is evaluated in section~\ref{sec:muon}. We summarize our findings in section~\ref{sec:summary}.

\section{Constraints on relevant MSSM parameters}
\label{sec:constraints}
In this section we revisit the present constraints on the MSSM parameters which will be crucial to our study. The most important parameters are the stau masses and mixing, which determine both the annihilation cross section $\sigma v (\neu \neu \rightarrow \tau^+ \tau^-)$ as well as the possible enhancement of $\Gamma(h\rightarrow \gamma \gamma)$. We also discuss the bounds on the Higgsino fraction of a 10~GeV neutralino, combined limits on the first two generation slepton masses from LEP and recent ATLAS results, and the LEP2 and recent LHC limits on charginos.
\subsection{LEP search for stau pair production}
\label{staumassbound}
The stau mass eigenstates, $\tilde{\tau}_{1,2}$, are linear combinations of the current eigenstates $\tilde{\tau}_{L,R}$. We take the masses $\mla$, $\mlb$ and the mixing angle  $\th$ as the free parameters in the stau sector. The relationship between the soft SUSY breaking parameters and the masses and mixing angle are reviewed in Appendix~\ref{sleptonmixing}. The most stringent bound on $\mla$ comes from the direct search of $\tla$ pair production at LEP,
\begin{figure}[htb!]
\centering
\includegraphics[scale=0.8]{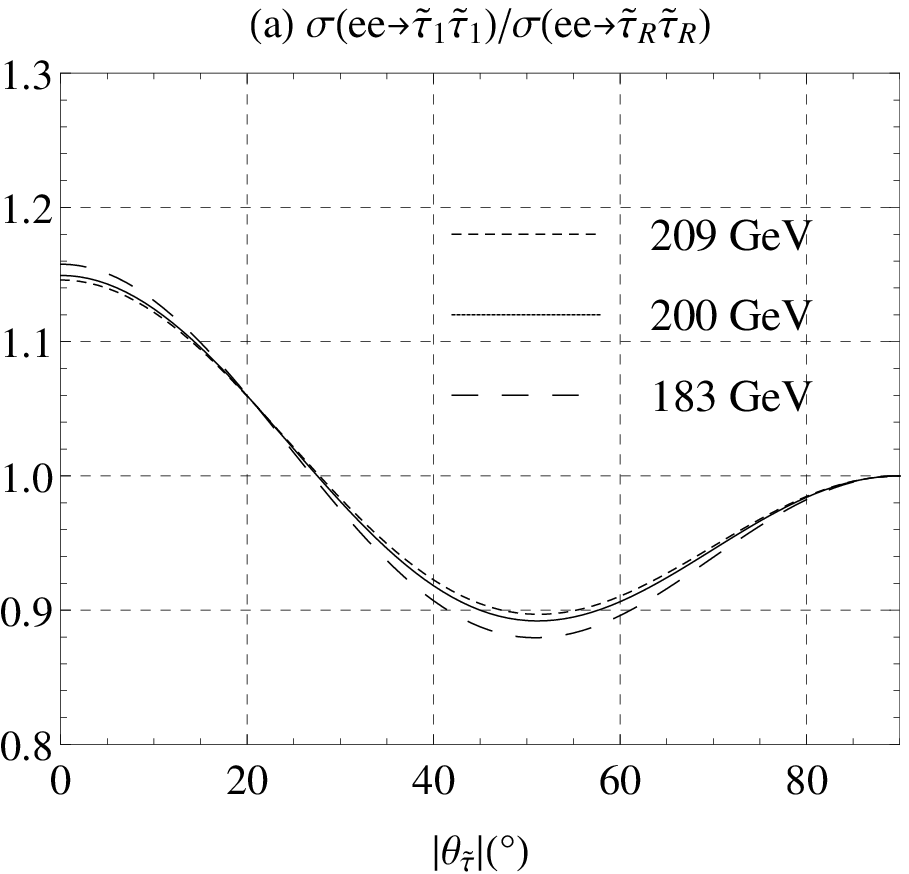}
\includegraphics[scale=0.79]{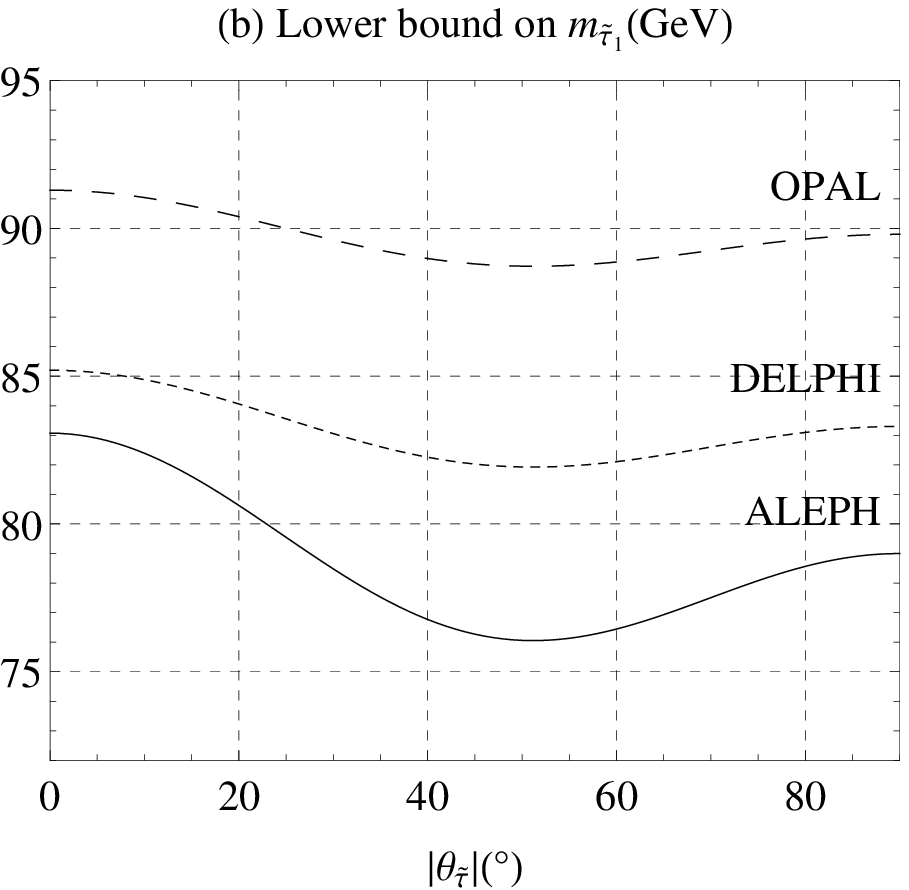}
\caption{\small {\it Left (a)} : Cross section for $e^+e^- \rightarrow \tilde{\tau}_1\overline{\tilde{\tau}_1}$, normalized to $\sigma(e^+e^-  \rightarrow \tilde{\tau}_R\overline{\tilde{\tau}_R})$, as a function of the stau mixing angle $|\th|$. The solid, dotted and dashed curves correspond to $\sqrt{s}=200, 209$ and $183$ GeV  respectively. {\it Right (b)} : The $95\%$ C.L. lower bound on the lightest stau mass as a function of $\th$ at $\sqrt{s}=200$ GeV. The ALEPH, DELPHI and OPAL bounds are shown by the solid, dotted and dashed curves respectively, where the regions above the curves are allowed.}
\label{plot1}
\end{figure}
where staus can be pair produced by the s-channel exchange of a $Z$ and a $\gamma$. Since the couplings of $\tilde{\tau}_{L,R}$ to the $Z$ boson are different, the stau pair production cross section depends on the stau mixing angle $\th$. The cross-section for $e^+e^- \rightarrow \tla \overline{\tla}$ is given by
\begin{align}
\sigma\bigl(e^-e^+\to \tilde{\tau}_1\overline{\tilde{\tau}_1}\bigr) = \frac{\beta^3s}{96\pi} \biggl[ & \Bigl|\frac{e^2}{s}
 + \frac{g_L}{s-m_Z^2}\Bigl( g_L\cos^2\th + g_R\sin^2\th \Bigr) \Bigr|^2 \nonumber \\
+
&\Bigl|\frac{e^2}{s} + \frac{g_R}{s-m_Z^2}\Bigl( g_L\cos^2\th + g_R\sin^2\th \Bigr) \Bigr|^2
\biggr],
\end{align}
where $s$ is the center-of-mass energy squared, $\beta = \sqrt{1-4m_{\tilde{\tau}_1}^2/s}$ is the stau velocity, and $g_{L,R}$ are given in terms of the $SU(2)_L$ gauge coupling and the weak mixing angle as $g_L = g_Z(-1/2+\sin^2\theta_w)$, $g_R = g_Z\sin^2\theta_w$, where $g_Z\cos{\theta_w}\sin{\theta_w}=g\sin{\theta_w}=e$. The cross section is symmetric about $\th=0^{\circ}$. In our convention for the mixing angle, $\tilde{\tau}_{1}$ is purely $\tilde{\tau}_{L}$ for $\th=0^{\circ}$, while it is purely $\tilde{\tau}_{R}$ for $|\th|=90^{\circ}$. In Figure~\ref{plot1} (a), we show the production cross section of $\tilde{\tau}_1\overline{\tilde{\tau}_1}$ normalized to that of $\tilde{\tau}_R\overline{\tilde{\tau}_R}$ as a function of the stau mixing angle. This ratio does not depend on $\mla$, but as can be seen from this figure, has a small dependence on $\sqrt{s}$. The solid, dotted and dashed curves denote the three LEP centre of mass energies of $\sqrt{s}=200, 209$ and $183$ GeV respectively. We can see from this figure that the cross section is maximum at $\th=0^{\circ}$ ($\tilde{\tau}_{1}=\tilde{\tau}_{L}$), while it reaches a minimum at around $|\th|\simeq50^{\circ}$, which corresponds to nearly maximal mixing between $\tilde{\tau}_{L}$ and $\tilde{\tau}_{R}$. 

In table \ref{tb1}, we summarize the $95\%$ C.L. lower limits on the lightest stau mass $\mla$ from the ALEPH, DELPHI and OPAL collaborations~\cite{stau1,stau2,stau4}.
These bounds are reported for both $|\th|=90^{\circ}$ and $|\th|=52^{\circ}$ cases,  under the assumptions ${\rm BR}(\tilde{\tau}_1\to\tau \tilde{\chi}_1^0)=1$ and $m_{\tilde{\chi}_1^0}=0$ (changing $m_{\tilde{\chi}_1^0}$ from $0$ to $10$ GeV increases the $\mla$ bound only by $\sim 1$ GeV). One of the implications of the assumption ${\rm BR}(\tilde{\tau}_1\to\tau \tilde{\chi}_1^0)=1$ is that the second lightest neutralino mass does not lie in between $\tilde{\tau}_1$ and $\tilde{\chi}_1^0$.\footnote{The analyses by L3 \cite{stau3} assume GUT unification conditions of the gaugino masses, and hence they do not give bounds for the limit ${\rm BR}(\tilde{\tau}_1\to\tau \tilde{\chi}_1^0)=1$.} For our numerical analysis, we quote both the ALEPH bound, which is the most conservative one, and the OPAL bound, which is the most stringent. Interestingly, the observed ALEPH bound is lower than the expected one, since they obtained a small excess over the background in acoplanar tau events in their 1999 data~\cite{ALEPH-excess}. In Figure~\ref{plot1} (b), we show the $95\%$ C.L. lower bound on $\mla$ as a function of $|\th|$, for $\sqrt{s}=200$ GeV by using the ALEPH, DELPHI and OPAL data.

\begin{table}[htb!]
\centering
\begin{tabular}{|c|c|c|c|}
\hline
  & ALEPH & DELPHI & OPAL\\
\hline
$|\th|=90^{\circ}$ & 79 & 83.3 & 89.8 \\
\hline
$|\th|=52^{\circ}$ &  76  & 81.9    & 88.7 \\
\hline
\end{tabular}
\caption{\small $95\%$ C.L. lower bound on the lightest stau mass (in GeV) by ALEPH, DELPHI and OPAL collaborations \cite{stau1,stau2,stau4}, assuming ${\rm BR}(\tilde{\tau}_1\to\tau \tilde{\chi}_1^0)=1$ and $m_{\tilde{\chi}_1^0}=0$.}
\label{tb1}
\end{table}
\subsection{Bounds on first two generation sleptons and chargino}
\label{slepton}

As far as the LEP2 bound on the lightest smuon mass is concerned, the most stringent limit comes from the OPAL collaboration~\cite{stau4}. Assuming that $BR(\tilde{\mu}_1\to\mu \tilde{\chi}_1^0)=1$ and $m_{\tilde{\chi}_1^0}=0$ GeV, the lower bound at $95\%$ C.L. is~\cite{stau4}
\begin{align}
m_{\tilde{\mu}_1} > 94.3\ \mathrm{GeV}\ \mathrm{at}\ |\theta_{\tilde{\mu}}|=90^{\circ},
\end{align}
where $|\theta_{\tilde{\mu}}|=90^{\circ}$ corresponds to the case  $\tilde{\mu}_1 = \tilde{\mu}_{R}$.  Recently, using the LHC data at 8 TeV, the ATLAS collaboration has reported the following bounds on the smuon mass at $95\%$ C.L.~\cite{LHCslepton}, using the same assumptions as above:
\begin{subequations}\label{smuonmassboundLHC}
\begin{align}
m_{\tilde{\mu}_1}& > 230,\ 94\ \mathrm{GeV} > m_{\tilde{\mu}_1}\ \mathrm{at}\ |\theta_{\tilde{\mu}}|=90^{\circ} ~(\tilde{\mu}_1 = \tilde{\mu}_{R}),\label{smuonbound1}\\
m_{\tilde{\mu}_1}& > 300,\ 94\ \mathrm{GeV} > m_{\tilde{\mu}_1}\ \mathrm{at}\ |\theta_{\tilde{\mu}}|=0^{\circ}~(\tilde{\mu}_1 = \tilde{\mu}_{L}).
\end{align}
\end{subequations}
The above bounds are obtained assuming that either $\tilde{\mu}_{R}$ or $\tilde{\mu}_{L}$ contributes to the cross-section. If both of them are light, common values for the left and right-handed slepton masses ($m_{\tilde{\ell}_{L}}=m_{\tilde{\ell}_{R}}$) between 90 GeV and 320 GeV are excluded by ATLAS at $95\%$ C.L. for a massless neutralino. The ATLAS bounds on the masses of $\tilde{e}_R$ and $\tilde{e}_L$ are the same as that for smuons in eq.(\ref{smuonmassboundLHC}), while the lower mass bound from LEP is $98$ GeV for $\tilde{e}_R$ with $m_{\neu}=10$ GeV (ALEPH~\cite{stau1}). The limits on the slepton masses from LEP and ATLAS are thus  complementary, excluding the whole mass region below 230 GeV for $\tilde{\ell}_{R}$ and 300 GeV for $\tilde{\ell}_{L}$.

The constraints on the chargino mass are especially important for the muon anomalous magnetic moment to be considered in section~\ref{sec:muon}, and also in determining the possible Higgsino fraction of a 10 GeV $\neu$. For the lower bound on the lighter chargino mass from LEP, the most stringent limit comes from DELPHI, and is given at $95\%$ C.L., by~\cite{stau2} 
\begin{align}
m_{\tilde{\chi}_1^{\pm}}& > 102.7\ \mathrm{GeV}\ \mathrm{for\ a\ Higgsino\ like}\  \tilde{\chi}_1^{\pm}, \nonumber \\
m_{\tilde{\chi}_1^{\pm}}& > 103.4\ \mathrm{GeV}\ \mathrm{for\ a\ wino\ like}\ \tilde{\chi}_1^{\pm},
\end{align}
which are obtained under the assumptions that the mass of the electron sneutrino $m_{\tilde{\nu}_e}>1000$ GeV and the mass difference between $\tilde{\chi}_1^{\pm}$ and $\tilde{\chi}_1^{0}$ is larger than $10$ GeV. Although when the chargino is wino like, t-channel exchange of the electron sneutrino can suppress the chargino pair production cross section, the dependence of the mass bound on $m_{\tilde{\nu}_e}$ is small. When $m_{\tilde{\nu}_e}=300$ GeV, the bound on the wino-like chargino mass reduces to 102.7 GeV.

In a scenario with a light stau, which is the focus of our study, a chargino can dominantly decay to final states containing a tau lepton, via an intermediate on-shell stau or a tau sneutrino. The ATLAS collaboration has searched for chargino pair production followed by the above decay chain, giving rise to a $2\tau^\pm + \met$ final state. Using the 8 TeV LHC data with an integrated luminosity of $20.7 \fb^{-1}$, the current ATLAS lower bound at $95\%$ C.L. is given by~\cite{ATLAS-chargino} 
\begin{equation}
m_{\tilde{\chi}_1^{\pm}} > 350\ \mathrm{GeV}\ \mathrm{for\ a\ wino\ like}\ \tilde{\chi}_1^{\pm}.
\end{equation}
The above bound is obtained in a simplified model assuming that the pair produced charginos give rise to the $2\tau^\pm + \met$ final state with 100\% branching fraction, and the lightest neutralino $\tilde{\chi}_1^0$ produced from the $\tilde{\chi}_1^{\pm}$ decay is massless. Thus combining the LEP and LHC limits, we shall consider charginos to be heavier than 350 GeV, although this lower bound gets relaxed once we depart from the above simplified model and consider wino-Higgsino mixed charginos. For example, we can translate the above bound for a wino-like  $\tilde{\chi}_1^{\pm}$ to ones for either a Higgsino-like or a maximally mixed wino-Higgsino  $\tilde{\chi}_1^{\pm}$ states as follows:

\begin{align}
m_{\tilde{\chi}_1^{\pm}}& \gtrsim 260\ \mathrm{GeV}\ \mathrm{for\ a\ Higgsino\ like}\  \tilde{\chi}_1^{\pm}, \nonumber \\
m_{\tilde{\chi}_1^{\pm}}& \gtrsim 300\ \mathrm{GeV}\ \mathrm{for\ a\ maximally\ mixed}\ \tilde{\chi}_1^{\pm}\ \bigl(|M_2/\mu| \simeq 1\bigr).
\end{align}

\subsection{Oblique corrections from staus}
\label{oblique}
In this section we study radiative effects of the staus to electroweak observables through universal gauge boson propagator corrections (oblique corrections). We find that the particle spectra relevant to this study are not only consistent with the precision data, but in some cases can improve the fit to the data. 

Before discussing the stau effect in detail, we briefly mention the effects from a very light $10$ GeV neutralino $\tilde{\chi}_1^0$ or a chargino satisfying the LEP and LHC bounds. As we shall see in section \ref{sec:zwidth}, a 10 GeV $\tilde{\chi}_1^0$ can only be an almost pure bino state. Since the bino couples neither to the $W$ boson nor to the $Z$ boson at the tree level, the oblique corrections from a $10$ GeV $\tilde{\chi}_1^0$ are thus negligible even with its very small mass. We have numerically verified this. As far as the chargino contribution is concerned, as discussed in section~\ref{slepton}, based on the combined LEP bound and the recent ATLAS bound in a scenario with a light stau, we take the lighter chargino to be heavier than 350 GeV. A chargino with this mass does not contribute significantly to the oblique corrections.

Following refs. \cite{EW2, EW3, EW1}, we adopt the three oblique parameters $S_Z$, $T_Z$ and $m_W$, and calculate the stau contribution to them. As discussed in Ref.~\cite{KHkyotoslide}, we take the following constraints on the shifts $\Delta S_Z$, $\Delta T_Z$ and $\Delta m_W$ from the reference SM values of $S_Z$, $T_Z$ and $m_W$ at the reference point, $(m_t, m^{}_{H_{SM}}, \Delta \alpha_{had}^{(5)}(m_Z^2))=(173.2\ \mathrm{GeV}, 125.5\ \mathrm{GeV}, 0.02763)$,
\begin{subequations}\label{fittingdata}
\begin{align}
\Delta S_Z + 2.23\Delta g_L^b& = 0.023 \pm 0.106,\label{ds}\\
\Delta T_Z - 0.47\Delta g_L^b& = 0.041 \pm 0.137,\label{dt}\\
\rho_{corr}&=0.91,
\end{align}
\end{subequations}
\begin{align}
\Delta m_W& = 0.025 \pm 0.015,
\end{align}
where $\rho_{corr}$ is the correlation between the errors in eqs. (\ref{ds}) and (\ref{dt}), and $\Delta g_L^b$ represents the shift of the $Zb_Lb_L$ vertex correction from the reference SM value, which has a significant $m_t$ dependence while all the other process specific (non-oblique) corrections do not. However the contributions from the MSSM particles to $\Delta g_L^b$ are negligible. The SM contributions to these shifts at different values of $m_t$ and $\Delta \alpha_{had}^{(5)}(m_Z^2)$ are shown in Ref. \cite{KHkyotoslide}.

\begin{figure}[htb!]
\centering
\includegraphics[scale=0.8]{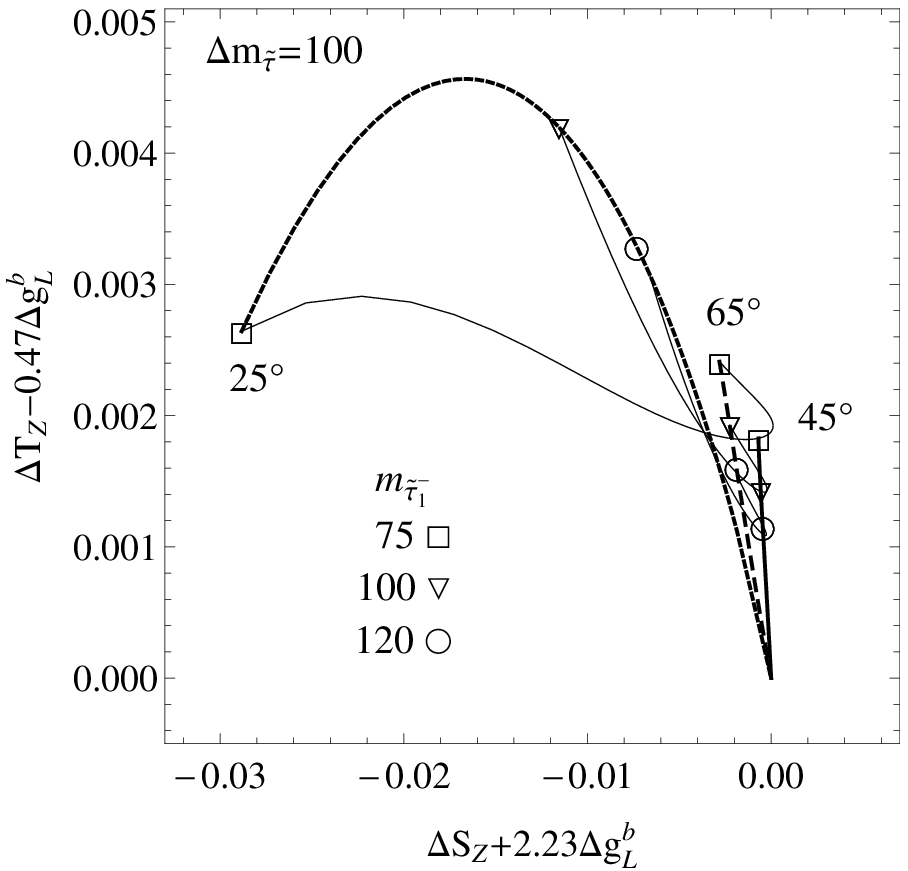}
\includegraphics[scale=0.82]{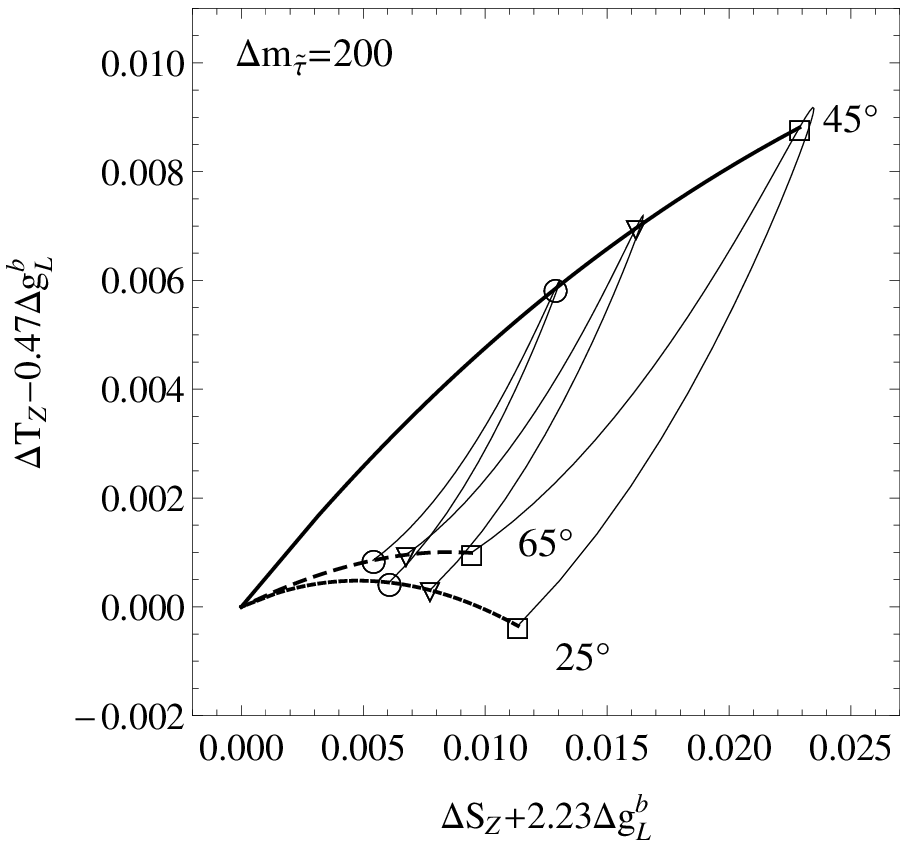}
\includegraphics[scale=0.78]{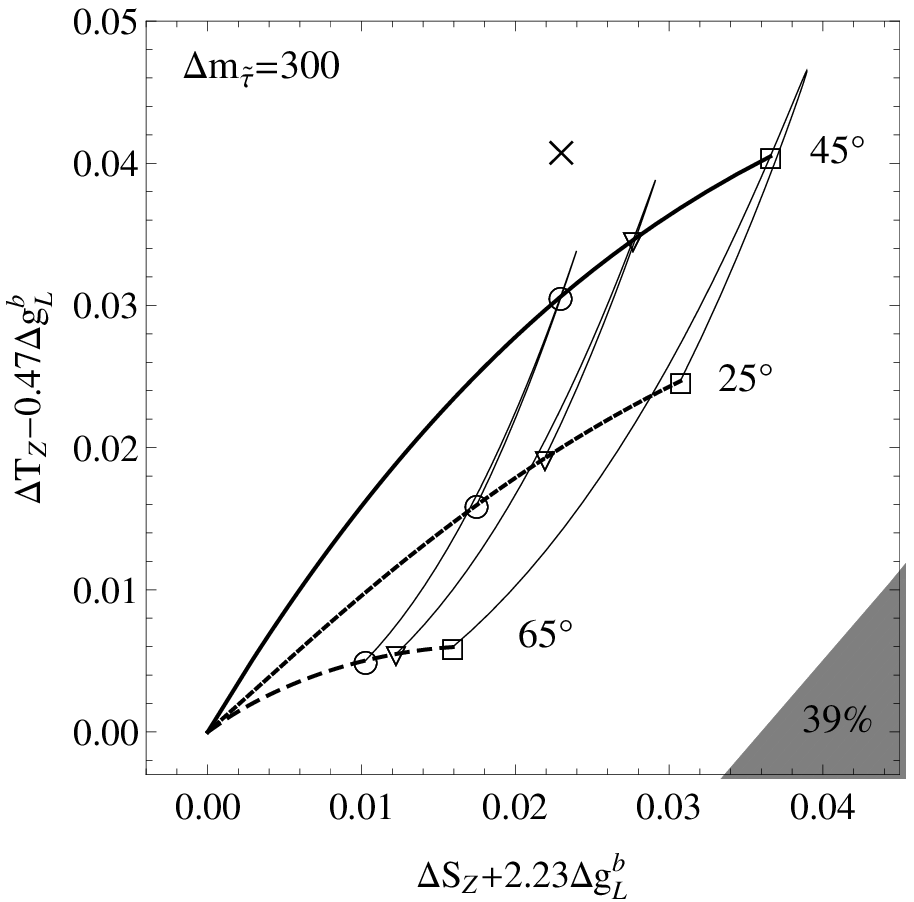}
\hspace{0.8cm}
\includegraphics[scale=0.72]{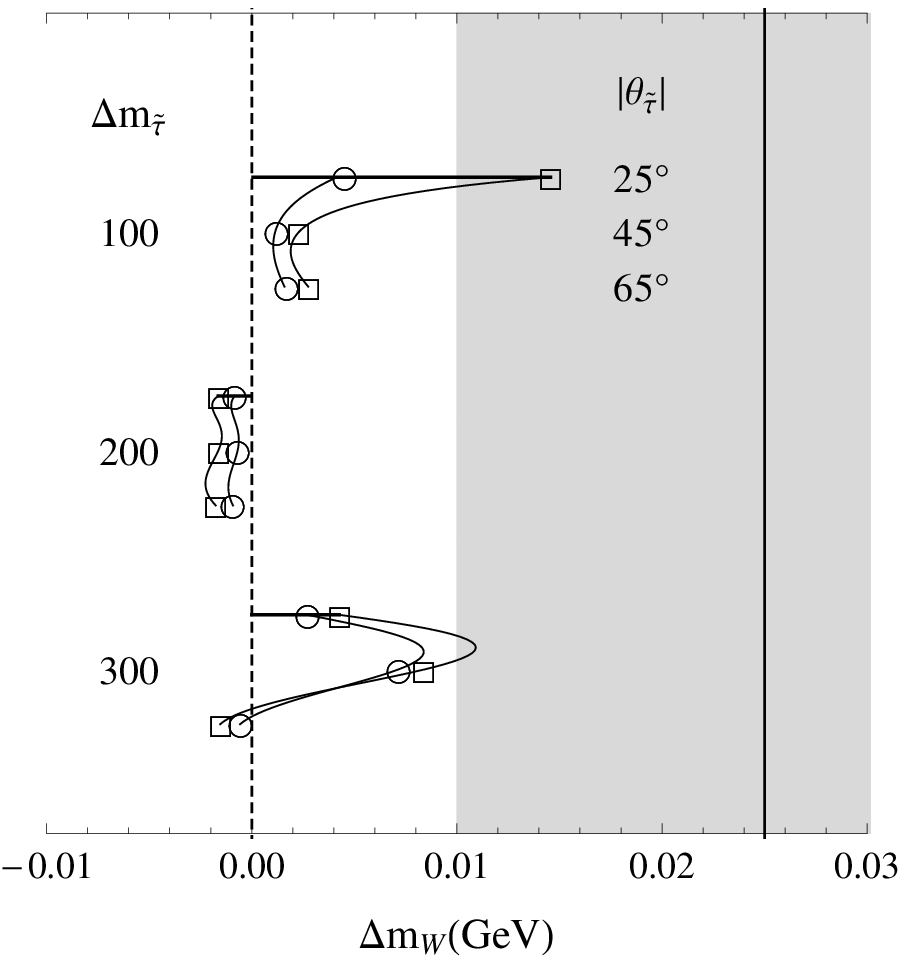}
\caption{\small The stau contribution to $\Delta S_Z$ and $\Delta T_Z$ in the upper two and the lower left plots and to $\Delta m_W$ in the lower right plot, with different values of $\Delta m_{\tilde{\tau}}=m_{\tilde{\tau}_2}-m_{\tilde{\tau}_1}$ and of the stau mixing angle $|\theta_{\tilde{\tau}}|$. The best-fit point is marked by $\times$ and the non-shaded region is consistent with the precision data at the $1\sigma$ ($39\%$ C.L.) in the ($\Delta S_Z$, $\Delta T_Z$) plane. In the $\Delta m_W$ plot, the SM prediction at the reference point is shown by the vertical dashed line at the origin, while the data is shown by the vertical solid line at $\Delta m_W=0.025$ GeV. The shaded region gives the $1\sigma$ ($67\%$ C.L.) uncertainty. For a detailed explanation of the plots, see text in section~\ref{oblique}.}
\label{plot9}
\end{figure}

In the upper two and the lower left plots of Figure~\ref{plot9}, we show the stau contribution to $\Delta S_Z$ and $\Delta T_Z$. The best-fit point is marked by a $\times$ and the $39\%$ confidence level contour given by the $1\sigma$ errors and the correlation of eq. (\ref{fittingdata}) are also shown. \footnote{Note that $\Delta \chi^2=1$ for 2 fitted parameters corresponds to the confidence level of $39\%$.} The non-shaded region is consistent with the precision data at the $39\%$ confidence level. We have taken $\tan\beta=10$, although the results in the figure do not change much at larger values of $\tan\beta$. The plots show the stau contribution for different values of the stau mass difference, $\Delta m_{\tilde{\tau}}=m_{\tilde{\tau}_2}-m_{\tilde{\tau}_1}=100$ GeV (upper left), $\Delta m_{\tilde{\tau}}=200$ GeV (upper right) and $\Delta m_{\tilde{\tau}}=300$ GeV (lower left). The open square $\Box$, triangle $\triangledown$ and circle $\circ$ indicate the radiative corrections for the lightest stau mass of $m_{\tilde{\tau}_1}=75, 100, 120$ GeV, respectively. The solid, dotted and dashed curves starting from the open squares denote the stau mixing angles $|\theta_{\tilde{\tau}}|=45^{\circ}, 25^{\circ}$ and $65^{\circ}$, respectively. All the trajectories converge to the reference SM point at the origin when $m_{\tilde{\tau}_1}$ gets large, as expected. The three thin curves connecting the open squares, triangles or circles show how the radiative corrections change as the stau mixing angle $|\theta_{\tilde{\tau}}|$ varies between $25^{\circ}$ and $60^{\circ}$. Note that the radiative corrections to $\Delta S_Z$ and $\Delta T_Z$ are symmetric about $\theta_{\tilde{\tau}}=0^{\circ}$. It is remarkable that the radiative corrections in the ($\Delta S_Z, \Delta T_Z$) plane grow as $\Delta m_{\tilde{\tau}}$ is increased from $100$ GeV (upper left) to $300$ GeV (lower left) when $m_{\tilde{\tau}_1}$ is fixed, and that the stau contribution improves the fit to the data when the mixing is nearly maximal $|\theta_{\tilde{\tau}}| \sim 45^{\circ}$. The effect is, however, not quantitatively significant, since the SM prediction at the origin is well inside of the $1\sigma$ ($39\%$ C.L.) allowed region. 

In the lower right plot of Figure \ref{plot9}, we show the stau contribution to the $W$ boson mass, $\Delta m_W$. The SM prediction at the reference point is shown by the vertical dashed line at the origin, while the data is shown by the vertical solid line at $\Delta m_W=0.025$ GeV. The shaded region shows the $1\sigma$ uncertainty. The stau contribution is expressed in the same manner as in the ($\Delta S_Z$, $\Delta T_Z$) plot, except that the $\triangledown$ points for $m_{\tilde{\tau}_1}=100$ GeV  are dropped in this plot.  It can be observed that the light stau makes the fit better than the SM especially when $\Delta m_{\tilde{\tau}}=100$ GeV for $|\theta_{\tilde{\tau}}|\sim 25^{\circ}$ or when $\Delta m_{\tilde{\tau}}=300$ GeV for $|\theta_{\tilde{\tau}}|\sim 45^{\circ}$. These improvements in the fit can be significant since the SM prediction is about $1.7 \sigma$ away from the data. 

To summarize our findings in this section, the mass spectra of the staus considered in this study do not contradict the electroweak precision data at the $1\sigma$ level. By combining the results in Figure \ref{plot9}, we find that the existence of a light stau satisfying $\Delta m_{\tilde{\tau}}=300$ GeV with nearly maximal stau mixing is slightly favored by the electroweak precision data. 

\subsection{Higgsino fraction of a $10$ GeV $\neu$}
\label{sec:zwidth}
Because of the bound on the chargino mass from LEP, a $\neu$ LSP of mass around $10$ GeV can neither be wino-like nor Higgsino-like. Furthermore, since the decay $Z\to \tilde{\chi}_1^0\tilde{\chi}_1^0$ is kinematically allowed in this mass range, and since only the Higgsinos  couple to the $Z$ boson at the tree level,  the Higgsino components of the neutralino LSP have an upper bound from the LEP limits on invisible decay width of the $Z$ boson. The $95\%$ C.L. upper bound on  the $Z$ boson invisible width, which does not originate from $Z\to\nu\bar{\nu}$, is given by~\cite{Zexp}
\begin{align}
\Gamma^Z_{\mathrm{inv}} < 2.0\ \mathrm{MeV}. 
\label{Z-LEP}
\end{align}
The partial width for the process $Z\to \tilde{\chi}_1^0\tilde{\chi}_1^0$ at the tree level is 
\begin{align}
\Gamma = \frac{G_Fm_Z^3}{12\sqrt{2}\pi}\Bigl( \bigl|(U^N_L)_{31}\bigr|^2 - \bigl|(U^N_L)_{41}\bigr|^2 \Bigr)^2
\biggl(1-\frac{4m_{\tilde{\chi}_1^0}^2}{m_Z^2}\biggr)^{3/2},
\label{Zwidth}
\end{align} 
where the neutralino mixing matrix $U_L^N$ is defined in Appendix~\ref{app1}. 
\begin{figure}[htb!]
\centering
\includegraphics[scale=0.8]{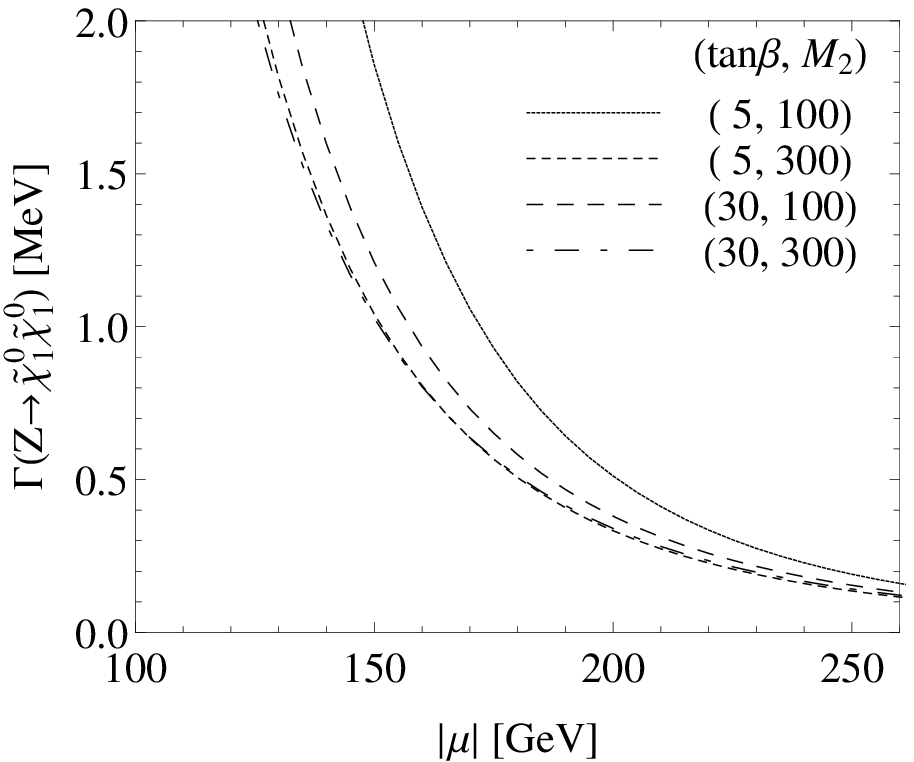}
\includegraphics[scale=0.845]{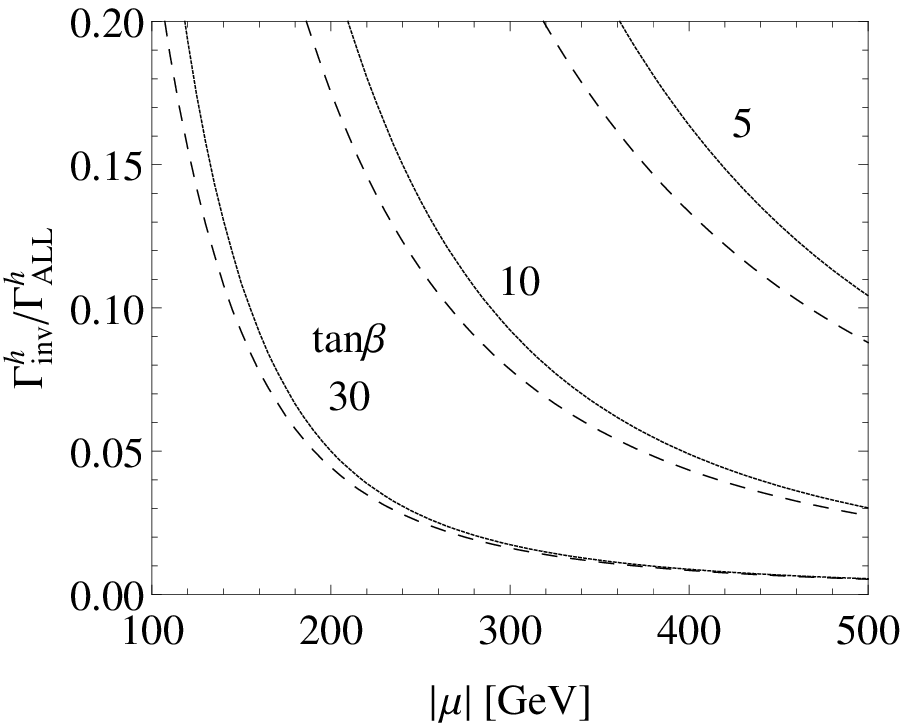}
\caption{\small Decay width of the process $Z\to \tilde{\chi}_1^0\tilde{\chi}_1^0$ (left panel) and the Higgs boson invisible branching ratio $\Gamma^H_{\mathrm{inv}}/\Gamma^H_{\mathrm{ALL}}$ (right panel), as a function of $|\mu|$, for different values of $M_2$ and $\tan\beta$, with $m_{\tilde{\chi}_1^0}$ fixed at $10$ GeV. $M_1 \mu$ and $M_2 \mu$ are chosen to be positive. In the right plot, the solid lines correspond to $M_2=100$ GeV and the dashed lines to $M_2=500$ GeV, with different values of $\tan{\beta}$ denoted near the lines.}
\label{plot5}
\end{figure}
The partial decay width grows as the Higgsino component in $\tilde{\chi}_1^0$ increases, giving rise to a lower limit on the $\mu$ parameter. In the left plot of Figure \ref{plot5}, we show $\Gamma(Z\to \tilde{\chi}_1^0\tilde{\chi}_1^0)$ as a function of $|\mu|$ for different values of $M_2$ and $\tan\beta$, keeping $m_{\tilde{\chi}_1^0}$ fixed at $10$ GeV. For this figure, we have used eq.(\ref{Zwidth}), with $m_Z=91.1876$ GeV, $\sin^2{\theta_w}=0.23116$ and $G_F=1.1664\times10^{-5}\ \mathrm{GeV}^{-2}$. As we can see from Figure~\ref{plot5}, the dependence of $\Gamma$ on the wino mass $M_2$ and $\tan\beta$ is small and it becomes smaller as $|\mu|$ becomes large compared to $m_Z$. We note in passing that the partial width $\Gamma$ in eq.~\ref{Zwidth} can vanish if $\bigl|(U^N_L)_{31}\bigr| = \bigl|(U^N_L)_{41}\bigr|$, which happens only for $\tan \beta=1$.

Next, we consider the contribution of a $10$ GeV $\tilde{\chi}_1^0$ to the Higgs boson invisible decay width, $\Gamma^h_{\mathrm{inv}}$. Although the present constraint on $\Gamma^h_{\mathrm{inv}}$ is weaker than that on $\Gamma^Z_{\mathrm{inv}}$, it nevertheless leads to a comparable upper bound on the Higgsino fraction of a $10$ GeV $\tilde{\chi}_1^0$. The Higgs boson couples to a mixed bino-Higgsino $\neu$ via the $U(1)_Y$ charge, with a coupling that depends on $\tan{\beta}$. The partial decay width of the CP-even Higgs boson for the process $h\to \tilde{\chi}_1^0 \tilde{\chi}_1^0$ at the tree level is 
\begin{align}
\Gamma = \frac{G_Fm_W^2m_h}{2\sqrt{2}\pi}
\biggl(1-\frac{4m_{\tilde{\chi}_1^0}^2}{m_h^2}\biggr)^{1/2}
\Biggl[
\bigl|U_{h\chi\chi}\bigr|^2 \biggl(1-\frac{2m_{\tilde{\chi}_1^0}^2}{m_h^2}\biggr)
- \mathrm{Re}\bigl[U_{h\chi\chi}^2\bigr]\frac{2m_{\tilde{\chi}_1^0}^2}{m_h^2}
\Biggr],
\label{Higgswidth}
\end{align}
where, for the lighter CP-even Higgs boson in the decoupling limit, the complex coupling $U_{h\chi\chi}$ is given, in terms of the neutralino mixing matrices $U_L^N$ and $U_R^N$ defined in Appendix~\ref{app1}, by
\begin{align}
U_{h\chi\chi} = \Bigl[ \bigl[(U^N_R)_{21}\bigr]^* (U^N_L)_{41} - \tan{\theta_w} \bigl[(U^N_R)_{11}\bigr]^* (U^N_L)_{41} \Bigr]&\times \sin{\beta}\nonumber \\
+ \Bigl[ \bigl[(U^N_R)_{21}\bigr]^* (U^N_L)_{31} - \tan{\theta_w} \bigl[(U^N_R)_{11}\bigr]^* (U^N_L)_{31} \Bigr]&\times \bigl(-\cos{\beta}\bigr).
\label{Hchichi}
\end{align}
Assuming that the Higgs boson production cross-section and partial decay widths to SM final states are the same as in the standard model, global fits to the current Higgs data from the LHC give the following upper bound on the Higgs boson invisible decay branching fraction at $95\%$ C.L.~\cite{Falkowski},
\begin{align}
\frac{\Gamma^h_{\mathrm{inv}}}{\Gamma^h_{\mathrm{ALL}}} \lesssim 0.2,
\end{align}
where $\Gamma^h_{\mathrm{ALL}}$ denotes the total decay width of the Higgs boson. In the present context, the  invisible branching fraction can be written as follows
\begin{align}
\frac{\Gamma^h_{\mathrm{inv}}}{\Gamma^h_{\mathrm{ALL}}}=\frac{\Gamma(h\to \tilde{\chi}_1^0 \tilde{\chi}_1^0)}{\Gamma^{H_{\mathrm{SM}}}_\mathrm{Total} + \Gamma(h\to \tilde{\chi}_1^0 \tilde{\chi}_1^0)}.
\end{align}
For our numerical analysis, we have used  $\Gamma^{H_{\mathrm{SM}}}_\mathrm{Total}=4.14$ MeV, with $m_h=125.5$ GeV~\cite{Higgswidth}. The right panel in Figure \ref{plot5} shows $\Gamma^h_{\mathrm{inv}}/\Gamma^h_{\mathrm{ALL}}$ as a function of $|\mu|$ for different values of $M_2$ and $\tan \beta$, using eq. (\ref{Higgswidth}). The solid lines correspond to $M_2=100$ GeV and the dashed lines to $M_2=500$ GeV, with different values of $\tan{\beta}$ denoted near the lines. As expected, the $M_2$ dependence is very small. In contrast to the process $Z\to \tilde{\chi}_1^0 \tilde{\chi}_1^0$, which nearly has a constant behaviour at moderate values of $\tan{\beta}$, $\Gamma(h\to \tilde{\chi}_1^0 \tilde{\chi}_1^0)$ shows a large $\tan{\beta}$ dependence. This is clearly seen from the leading term in $|U_{h\chi\chi}|$,
\begin{align}
|U_{h\chi\chi}| = \tan^2{\theta_w}m_W\frac{\sin{2\beta}}{|\mu|}+O(m_Z^2/\mu^2).
\end{align}
\\

Thus, from the above considerations, we see that a scenario with a $10$ GeV $\tilde{\chi}_1^0$ in the MSSM with a heavy Higgsino and/or large values of $\tan{\beta}$ is compatible with the constraints on $\Gamma^Z_{\mathrm{inv}}$ and $\Gamma^h_{\mathrm{inv}}$. We have seen in  section \ref{slepton} that the ATLAS results in the $2\tau^\pm+\met$ channel, assuming the presence of a light stau, lead to a lower bound on the lighter chargino mass of $m_{\tilde{\chi}^{\pm}_1}\gtrsim 350$ GeV for a wino-like $\tilde{\chi}^{\pm}_1$.  Although this LHC constraint can become weaker if the lighter chargino is a pure Higgsino state, henceforth, we shall conservatively consider $|\mu|>350$ GeV. For such values of $|\mu|$, the Higgsino fraction of a 10 GeV $\neu$ is smaller than $1.5\%$.

\section{Light $\neu$ DM annihilation to tau leptons}
\label{sec:annihilation}
In this section we consider the annihilation process $\neu \neu \rightarrow \tau^+ \tau^-$ in the limit of zero relative velocity $v$ between the $\neu$'s (the average velocity of DM particles in the galactic halo is estimated to be $v \sim 10^{-3}c$). Light $\neu$ DM can annihilate to lighter standard model (SM) fermions via $t$ and $u$ channel sfermion exchange, and $s$-channel exchange of either a $Z$ boson or one of the three neutral Higgs bosons in the MSSM. As we have seen in the previous section, constraints on the chargino mass and the $Z$ and Higgs boson invisible widths dictate that a $\mathcal{O}(10 \gev)$ neutralino has to be dominantly composed of a bino ($\B$).  Therefore, in this section we first discuss the pure bino approximation, and then comment on the modifications due to a possible small Higgsino fraction.
\subsection{Pure bino annihilation}
Since a pure bino couples  neither to the $Z$ boson nor to the Higgs bosons, only the sfermion exchange diagrams contribute to their pair annihilation. It is well-known that in the absence of any mixing between the sfermion gauge eigenstates $\tilde{f}_L$ and $\tilde{f}_R$, the s-wave component of the velocity averaged annihilation rate $\langle \sigma v \rangle$ to SM fermions is proportional to $m_f^2$, where $m_f$ is the corresponding fermion mass. In the presence of $\tilde{f}_L- \tilde{f}_R$ mixing, which can be large for third generation sfermions, there is an additional contribution to the annihilation amplitudes proportional to $\mB$. Combining these two facts, since only the s-wave contribution is relevant to the annihilation of DM particles in the present universe, for a 10 GeV $\B$, the important modes to consider are $\tau^+ \tau^-$ and $b \bar{b}$. Here, we focus on the $\tau^+\tau^-$ mode and determine the parameters in the stau sector which can give $\langle \sigma v \rangle={\langle \sigma v\rangle}_0$, where the value of the annihilation rate required to fit the Fermi Bubble data is estimated to be~\cite{HS-1} 
\begin{equation}
{\langle \sigma v\rangle}_0=2 \times 10^{-27} {\rm cm}^3/{\rm s}.
\label{eqn-value}
\end{equation}
Since a light stau with $\mla \gtrsim 90 \gev$ is allowed by the LEP data, the annihilation cross-section for  $\B \B \rightarrow \tau^+ \tau^-$  can be significant. The relevant s-wave annihilation cross-section, in the limit of zero relative velocity $v$ between the neutralinos, is given by~\cite{Jungman,Drees,Falk,Ellis}
\begin{align}
\langle \sigma v \rangle \xrightarrow{v \rightarrow 0} \frac{{g^\prime}^4\beta_\tau}{128 \pi} \biggl[ & \frac{4 \left(\mt \YR^2 s^2 +2\mB \YL \YR s c + \mt \YL^2 c^2  \right)}{\Delta_1} \nonumber \\
 & +\frac{4 \left(\mt \YR^2 c^2 -2\mB \YL \YR s c + \mt \YL^2 s^2  \right)}{\Delta_2} \biggr]^2 ,
\end{align}
where, $g^\prime=e/\cos\theta_w$ is the $U(1)_{\rm Y}$ gauge coupling, $\YL=-1/2$ and $\YR=-1$ are the hypercharges of $\tau_L$ and $\tau_R$, $\beta_\tau=\sqrt{1-\mt^2/\mB^2}$ and $\Delta_i=\mli^2+\mB^2-\mt^2$. We have used the shorthand $s=\st$ and $c=\ct$,  with $-\pi/2 < \th < \pi/2$ (our notation for the slepton masses and mixing are summarized in Appendix~\ref{sleptonmixing}).

\begin{figure}[htb!]
\centering
\includegraphics[scale=0.65]{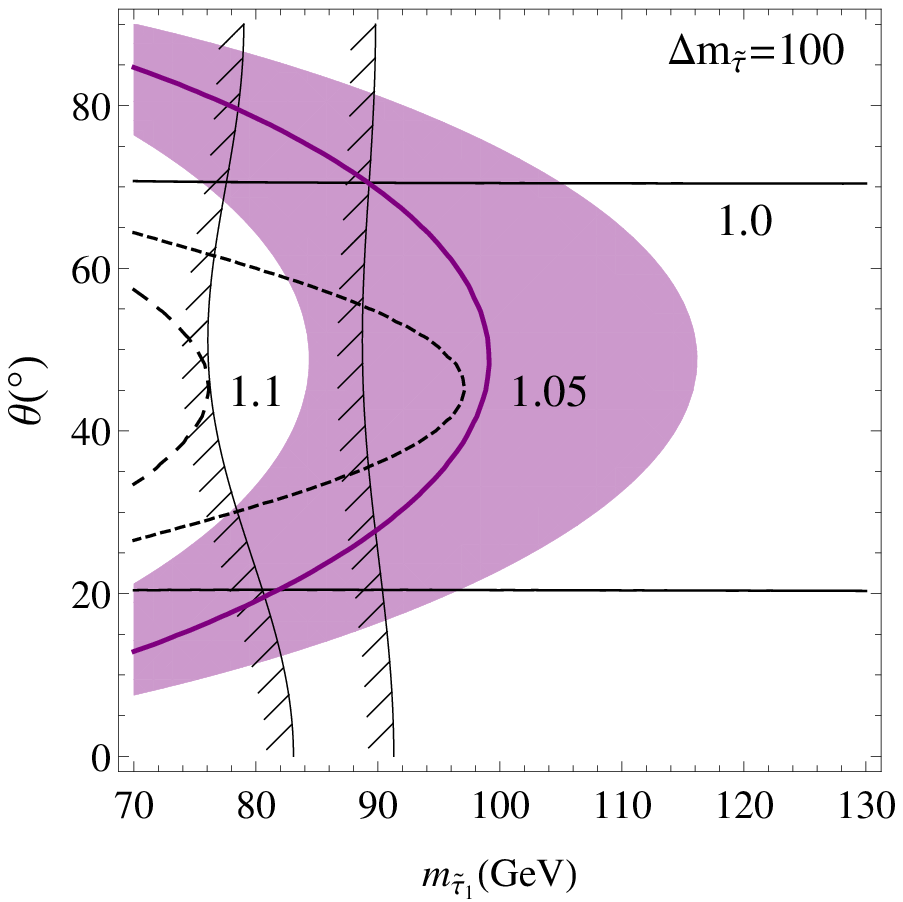}
\includegraphics[scale=0.67]{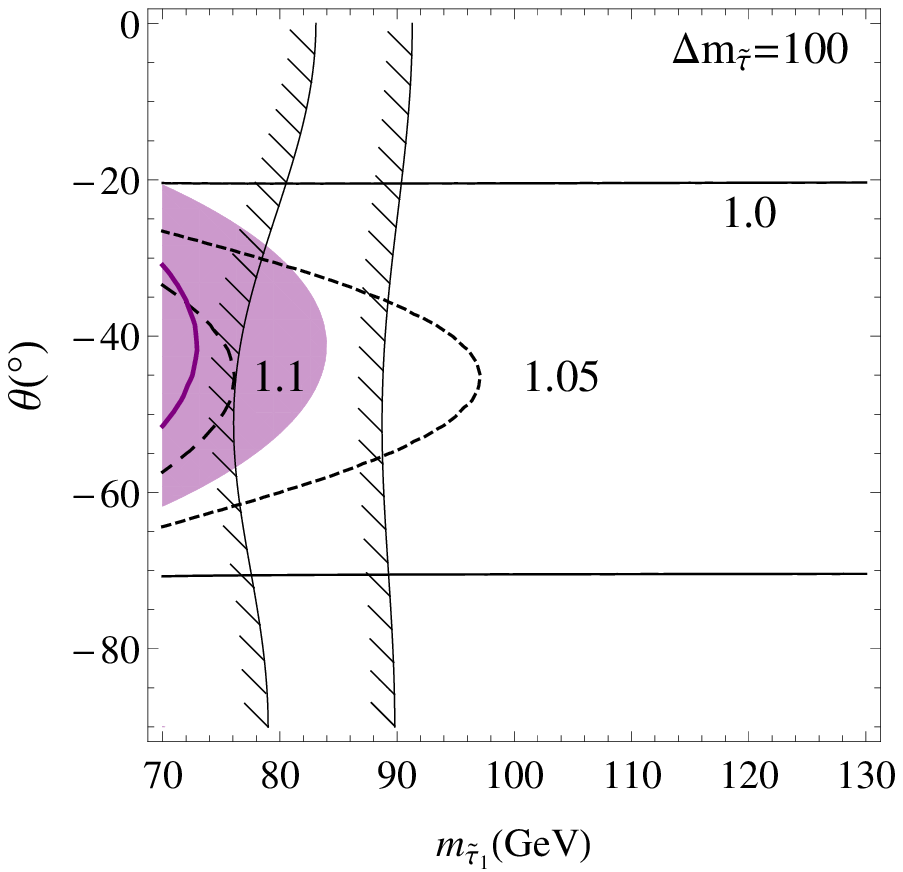}
\includegraphics[scale=0.65]{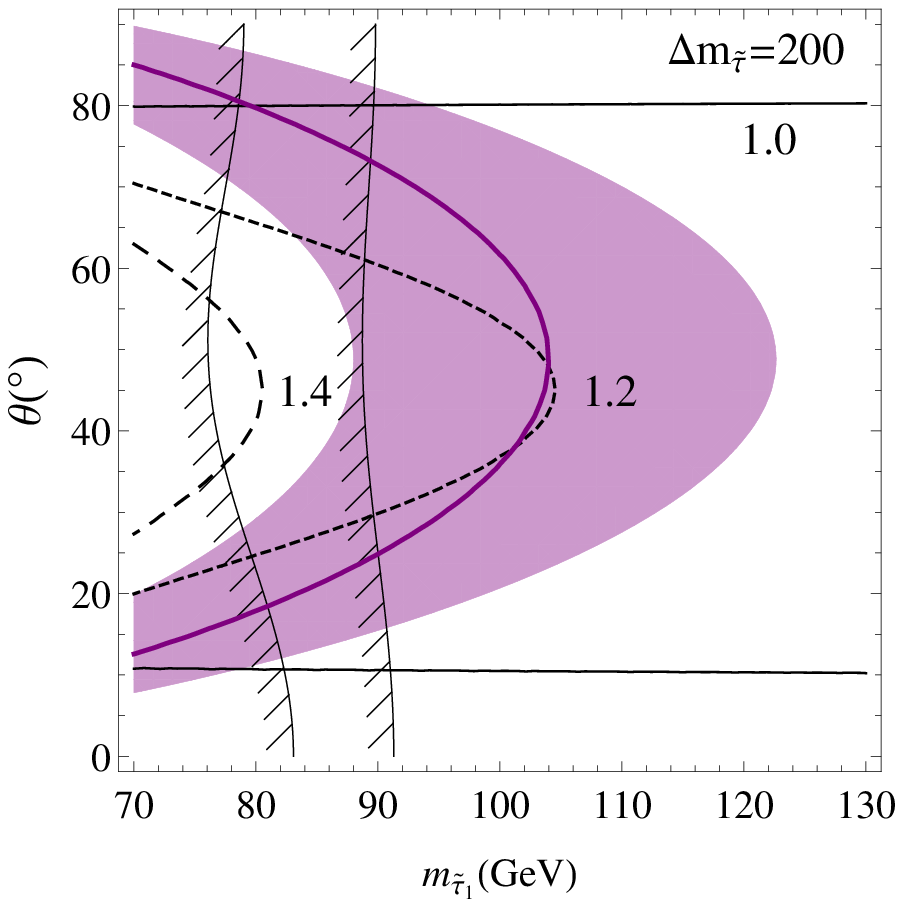}
\includegraphics[scale=0.67]{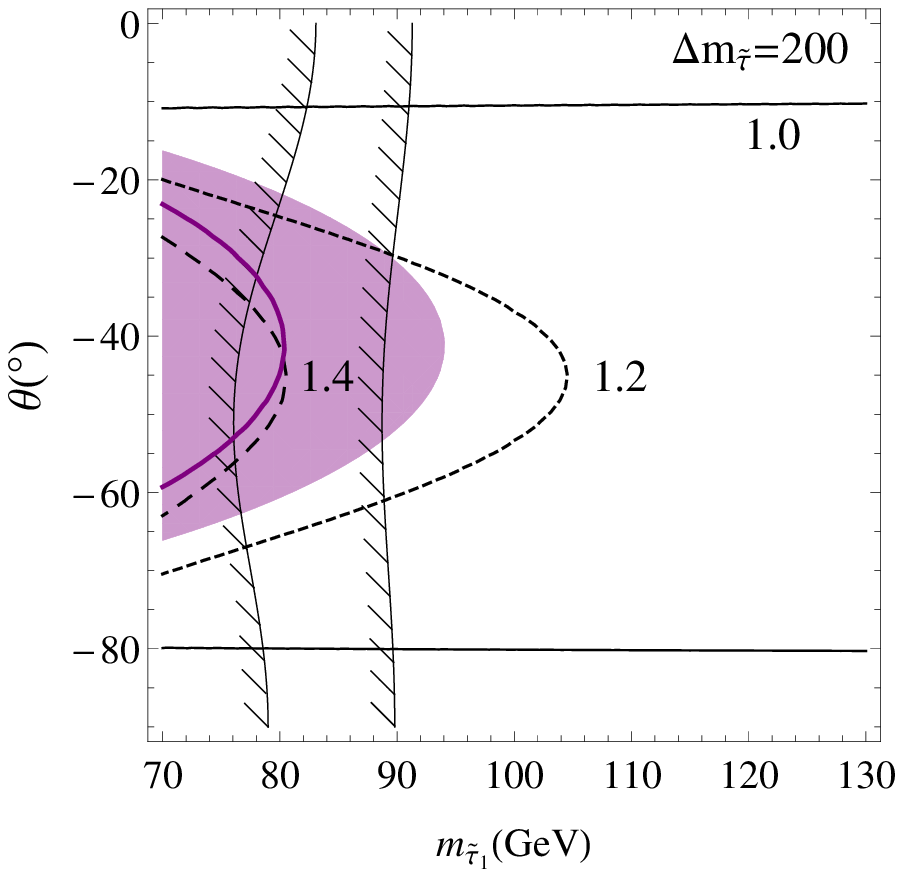}
\includegraphics[scale=0.65]{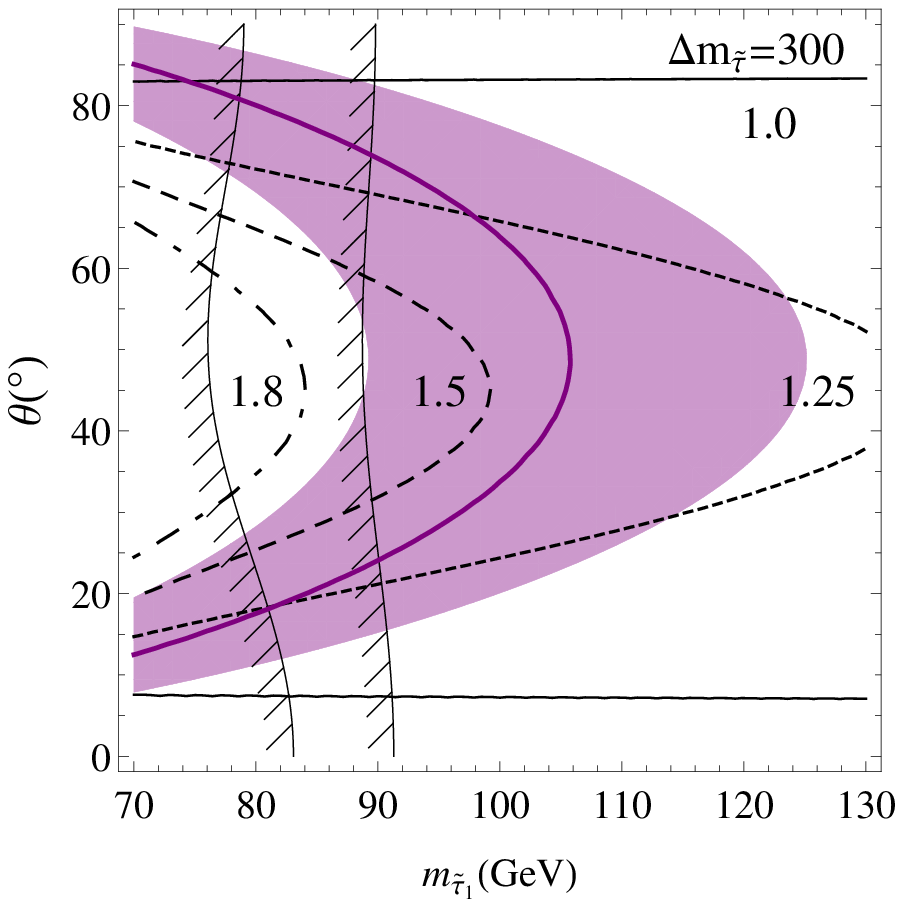}
\includegraphics[scale=0.67]{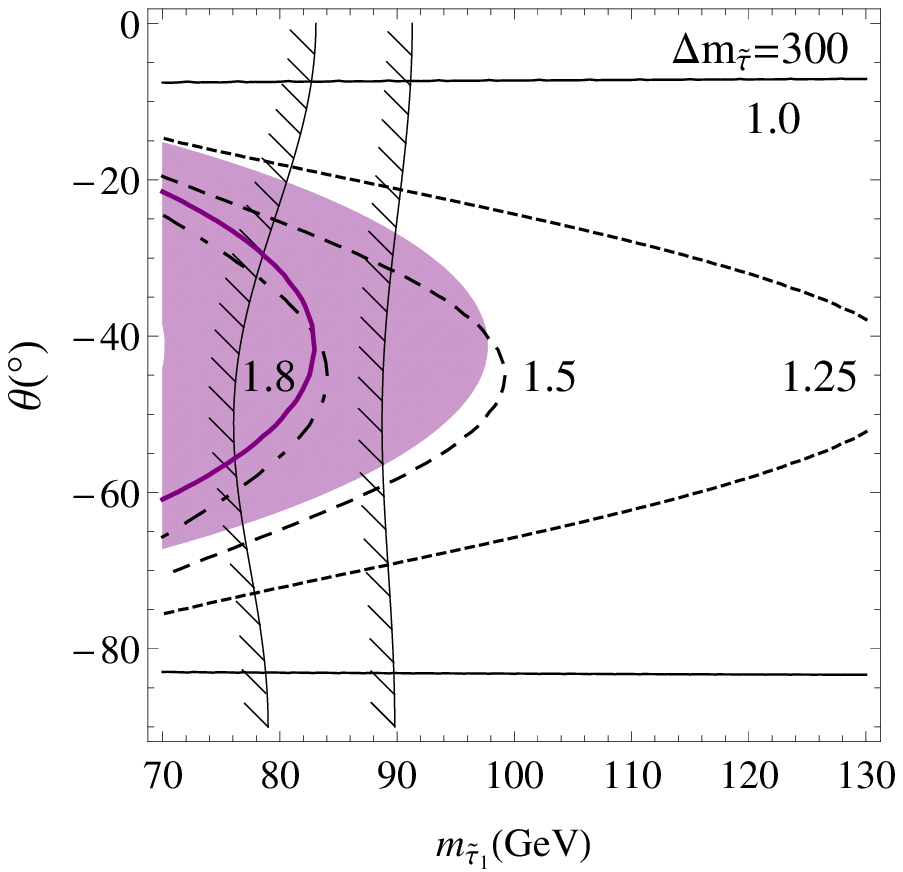}
\caption{\small Contours in the $\mla-\th$ plane for $\langle \sigma v \rangle=\langle \sigma v\rangle_0$, with $\langle \sigma v \rangle_0=2 \times 10^{-27} {\rm cm}^3/{\rm s}$ (solid purple curve) and the ratio of the two photon decay width of Higgs, $\Gamma(h\to \gamma\gamma)/\Gamma(H_{\rm SM} \to \gamma\gamma)$ (black solid, dotted and dashed curves), with different values of the mass difference, $\Delta m_{\tilde{\tau}}=m_{\tilde{\tau}_2}-m_{\tilde{\tau}_1} = 100$ (upper), $200$ (middle) and $300$ GeV (lower). The shaded region gives $\langle \sigma v \rangle_0/2 < \langle \sigma v \rangle< 2\langle \sigma v\rangle_0$. The vertical curves show the lower bounds on $\mla$ as a function of $\th$ coming from LEP.}
\label{plotdm1}
\end{figure}

Since the terms proportional to $\mB$ coming from $\tla$ and $\tlb$ exchange diagrams have opposite signs in the amplitude, for fixed values of $\mla$ and $\th$, $\langle \sigma v \rangle$ increases with increasing $\mlb$, and eventually reaches an asymptotic value. The mixing-induced term is maximized for $\th=\pi/4$. In Figure~\ref{plotdm1}, we show purple solid contours in the $\mla-\th$ plane, for $\Delta m_{\tilde{\tau}} = \mlb-\mla=100, 200$ and $300$ GeV, satisfying $\langle \sigma v \rangle=\langle \sigma v \rangle_0$. The vertical curves show the lower bounds on $\mla$ as a function of $\th$ coming from LEP, where both the conservative ALEPH~\cite{stau1} and most stringent OPAL~\cite{stau4} bounds of Figure~\ref{plot1} are given. As discussed in Refs.~\cite{HS-1,Gordon}, since there are possible systematic uncertainties in the value of $\langle \sigma v \rangle_0$ coming from the subtraction of the inverse Compton scattering component, we show by shaded area a region corresponding to $\langle \sigma v \rangle_0/2<\langle \sigma v \rangle<2  \langle \sigma v \rangle_0$. As clearly shown by the figures, the annihilation rate $\langle \sigma v \rangle_0$ to $\tau^+\tau^-$ can be obtained in a certain region of the $\mla-\th$ parameter space satisfying the LEP constraints. For example, the required value of $\mla$ for the case of maximal mixing ($\th=\pi/4$) is 103.5 GeV, for $\mlb=300 \gev$, according to the left plot in the second row of Figure~\ref{plotdm1}. The importance of the large stau mixing in obtaining the necessary $\langle \sigma v \rangle$ is clearly seen in the figures. For $\th>0$, the range of the mixing angle that can accommodate $\langle \sigma v \rangle = \langle \sigma v \rangle_0$, with $\tla$ mass above the LEP limit is $30^\circ < \th < 70^\circ$ for $\Delta m_{\tilde{\tau}} = 100$ GeV and $25^\circ < \th < 75^\circ$ for $\Delta m_{\tilde{\tau}} = 200,300$ GeV. Thus beyond around $\Delta m_{\tilde{\tau}} =200$ GeV, the asymptotic value of $\langle \sigma v \rangle$ is reached for a given $\mla$ and $\th$, and a further increase of the stau mass splitting has no significant effect. In all these cases, $\tla$ is required to be lighter than about 106 GeV for $\langle \sigma v \rangle = \langle \sigma v \rangle_0$, and it is relaxed to 125 GeV if we allow $\langle \sigma v \rangle=\langle \sigma v \rangle_0/2$. On the other hand, for the case of $\th <0$, shown in the right column of Figure~\ref{plotdm1}, it is significantly more difficult to obtain a sufficiently large annihilation rate with a $\tla$ that satisfies the LEP bound. Even at the maximal mixing of $\th=-\pi/4$, $\Delta m_{\tilde{\tau}} > 200$ GeV is needed to obtain $\langle \sigma v \rangle > \langle \sigma v \rangle_0/2$, for $\mla >90$ GeV. We have checked all our results for the annihilation rate using {\tt micrOMEGAs-3.2}~\cite{micromega}, with the MSSM mass spectrum from the soft SUSY-breaking parameters generated by the code {\tt SuSpect-2.3}~\cite{suspect}~\footnote{The convention for stau mixing matrix we have used (see Appendix~\ref{sleptonmixing}) is different from the one used in {\tt SuSpect}~\cite{suspect}, and one has to translate $\th$ accordingly for comparison.}.

Since the $\B \B \rightarrow \tau^+ \tau^-$ annihilation cross-section is sensitive to the sign of the mixing angle $\th$, we review our notation for the slepton mixing matrix and the MSSM parameters in Appendix~\ref{sleptonmixing}. The positive mixing angle favoured by the required annihilation rate in Figure~\ref{plotdm1}, corresponds to the condition
\begin{equation}
m_{\tilde{\tau}_{LR}}^2 = m_{\tau} (A_{\tau} - \mu \tan \beta) >0
\label{off-d}
\end{equation}
in the MSSM, with the octant $0<\th<\pi/4$ ($\pi/4 <\th <\pi/2 $) for $m_{\tilde{\tau}_{R}}^2 >m_{\tilde{\tau}_{L}}^2$ ($m_{\tilde{\tau}_{R}}^2 <m_{\tilde{\tau}_{L}}^2$), according to eqs.(\ref{mLR}) and (\ref{octant}). This implies $\mu<0$ for large $\tan \beta$ or a very large positive value of $A_\tau$. It should be noted, however, that if $m_{\tilde{\tau}_{L}}^2 \approx m_{\tilde{\tau}_{R}}^2$ holds in the mass-squared matrix (\ref{massmatrix}), the relatively small magnitude of the off-diagonal element (\ref{off-d}) suffices to give the maximal mixing.

The magnitude of the stau left-right (L-R) mixing may also be constrained if we impose the (meta-) stability condition of the electroweak vacuum. In the context of previous studies on the enhancement of $\Gamma_{h \rightarrow \gamma \gamma}$ by a light $\tilde{\tau}_1$ and large L-R mixing in the stau sector, the consequences of this L-R mixing in de-stabilizing the electroweak vacuum was investigated~\cite{vacuum,Kitahara}. It is possible, in this context, to have a charge-breaking minimum of the effective scalar potential which is deeper than the electroweak symmetry breaking minimum. Quantum tunneling effects can cause vacuum decay to the true minimum of the potential, and if the decay time is larger than the age of the universe, the electroweak vacuum will be meta-stable. Based on the calculation of the upper bound on $|\mu \tan\beta|$ from the requirement of meta-stability using semi-classical methods, a fitting formula was obtained in Ref.~\cite{Kitahara} which we use to calculate this upper bound as a function of the stau soft-masses $m_{\tilde{\tau}_{L}}$ and $m_{\tilde{\tau}_{R}}$. As an example, for equal left and right soft masses, we can obtain a maximal mixing, $\theta_{\tilde{\tau}}=45^\circ$ and the stau mass splitting is determined by $\mu \tan \beta$ (for $A_\tau=0$). With $m_{\tilde{\tau}_1}=100$ GeV, in order to match the annihilation rate $\sigma v=\sigma v_0$, we would require $\Delta m_{\tilde{\tau}}=115$ GeV. Such a spectrum can be obtained if $\mu \tan \beta=10.3$ TeV, which is compatible with the vacuum meta-stability condition, which demands, in this case,   $|\mu \tan\beta|<16.7$ TeV. For the higher mass splitting cases considered in Figure~\ref{plotdm1}, if we fix $\mla=100$ GeV and $\th=45^\circ$ for simplicity (both of which are consistent with the annihilation rate requirement), then we find that in order to obtain $\Delta m_{\tilde{\tau}}=200$ and 300 GeV, we would require $\mu \tan \beta = 22.6$ and 42.4 TeV respectively. The vacuum meta-stability condition for the above two parameter points leads to $|\mu \tan \beta| < 27$ TeV and 39 TeV respectively, and therefore, having a mass splitting as large as 300 GeV might not be possible if we impose this condition. However, as we have seen above, the neutralino annihilation rate already reaches an asymptotic value at $\Delta m_{\tilde{\tau}}=200$ GeV, and therefore, the vacuum meta-stability requirement does not pose any significant constraint in our scenario.

We note in passing that the value of the s-wave annihilation rate required to fit the Fermi Bubble data, eq.(\ref{eqn-value}), is consistent with the bounds on $\langle \sigma v \rangle$ obtained by the Fermi collaboration, based on non-observation of significant gamma-ray excess from dwarf spheroidal galaxies~\cite{Ackermann}. For a 10 GeV DM annihilating to $\tau^+ \tau^-$, the current Fermi-LAT data excludes $\langle \sigma v \rangle > 1.5 \times 10^{-26} {\rm cm}^3/{\rm s}$ at $95 \%$ C.L.~\cite{Ackermann}. 

\subsection{Implications of a small Higgsino fraction}
\begin{table}[htb!]
\centering
\begin{tabular}{|c|c|c|c|}
\hline
$|\mu|$  & Gaugino Fraction & Higgsino Fraction & $\langle \sigma v \rangle (\neu \neu \rightarrow \tau^+ \tau^-)_{v\rightarrow 0}$ \\
(GeV)   &   &    &  ($10^{-27} {\rm cm}^3$/s)\\
\hline
350 &0.984  &0.015  & 2.28 \\
500 &0.992  &0.008  & 2.20 \\
800 &0.998  &0.003  & 2.16  \\
\hline
\end{tabular}
\caption{\small Variation of $\langle \sigma v \rangle (\neu \neu \rightarrow \tau^+ \tau^-)_{v\rightarrow 0}$ with $|\mu|$, and hence the gaugino-Higgsino fractions of a 10 GeV $\neu$, with all other relevant parameters fixed as: $\mla=100$ GeV, $\mlb=215$ GeV, $\th=45^{\circ}$, $m_{\tilde{\chi}_1^+}=350$ GeV, $m_h = 126$ GeV, $m_A=500$ GeV and $\tan \beta=30$.}
\label{muVary}
\end{table}

So far, in discussing the annihilation rate $\langle \sigma v \rangle (\neu \neu \rightarrow \tau^+ \tau^-)$, we have assumed that $\neu$ is a pure bino. In section~\ref{sec:zwidth}, we have studied the invisible width constraints of the $Z$ boson (at LEP) and the Higgs boson (at LHC) decaying into a $\neu \neu$ pair. These constraints restrict the value of the $\mu$ parameter depending upon the choice of $M_2$ and $\tan \beta$, and therefore imply an upper bound on the Higgsino fraction of a $10$ GeV $\neu$. In addition, the recent chargino mass bound from LHC in presence of a light stau puts stronger constraints on $|\mu|$, as seen in section~\ref{slepton}. We take the minimum value of $|\mu|=350$ GeV, making a conservative estimate based on the chargino mass constraint from the LHC and briefly discuss the implications of this small Higgsino fraction in $\neu$. We define the Higgsino fraction of the lightest neutralino $\neu$ as $|N_{13}|^2+|N_{14}|^2$ from the expansion 
\begin{equation}
\tilde{\chi}_{1L}^0= N_{11} \tilde{B}_L+N_{12}\tilde{W}^3_L+N_{13} \tilde{H}^0_{dL}+N_{14} \tilde{H}^0_{uL}
\end{equation}
defined in eq.(\ref{mass-eigen}) of Appendix~\ref{app1}, and accordingly the gaugino fraction as $|N_{11}|^2+|N_{12}|^2$.

 In table~\ref{muVary}, we show the effect of varying $|\mu|$, and hence the gaugino-Higgsino fractions of a $10$ GeV $\neu$, on $\langle \sigma v \rangle (\neu \neu \rightarrow \tau^+ \tau^-)_{v\rightarrow 0}$ (the cross-sections have been obtained using {\tt micrOMEGAs-3.2}). All other relevant parameters have been fixed as: $\mla=100$ GeV, $\mlb=215$ GeV, $\th=45^{\circ}$, $m_{\tilde{\chi}_1^+}=350$ GeV, $m_h = 126$ GeV, $m_A=500$ GeV and $\tan \beta=30$. The value of $\tan \beta$ is chosen to be rather large, since as we shall see in section~\ref{sec:muon}, in order to explain the muon $g-2$ anomaly within $1\sigma$ with a chargino and smuon mass spectrum consistent with the recent ATLAS bounds, $\tan \beta$ is required to be greater than 20. The stau mass splitting is kept fixed while varying $|\mu|$ by choosing appropriate values of the trilinear scalar soft term $A_\tau$. As we can see from this table, the Higgsino fraction is rather small for $|\mu| =350$ GeV (1.5\%), and as the Higgsino fraction decreases, the annihilation rate to $\tau^+ \tau^-$ also decreases for this choice of $\tan \beta$. There is also a $\tan \beta$ dependence of $\langle \sigma v \rangle$ because the Higgsino components ($N_{13} \tilde{H}^0_{dL}$ and $N_{13}^* \tilde{H}^0_{dR}$) couple to $\tli-\tau$  with the Yukawa coupling, $y_{\tau}=\sqrt{2}m_{\tau}/(v \cos \beta)$, which grows with $\tan \beta$ (the bino component of $\neu$ ($N_{11}\B_{L}$ and $N_{11}^*\B_{R}$) couples to $\tli-\tau$ with the $U(1)_{\rm Y}$ gauge coupling of $g^\prime=e/\cos\theta_w \approx 0.355$). For $\tan \beta =10$, the Yukawa coupling $y_{\tau}\approx 0.103$ is significantly smaller than the gauge coupling, while for $\tan \beta=30$ they become comparable ($y_{\tau}\approx 0.307$). In the above parameter choices, if we just change $\tan \beta$ from 30 to 10, keeping the other parameters fixed, for $|\mu|=500$, although the gaugino and Higgsino fractions remain almost the same (0.992 and 0.008 respectively), $\langle \sigma v \rangle (\neu \neu \rightarrow \tau^+ \tau^-)_{v\rightarrow 0}$ now becomes $ 1.95 \times 10^{-27} {\rm cm}^3$/s. Therefore, in order to match the neutralino annihilation rate with the Fermi Bubble data, having a small Higgsino fraction for $\tilde{\chi}^0_1$ and a relatively large $\tan \beta$ will be helpful, since it opens up a parameter region in the stau sector with a somewhat larger $m_{\tilde{\tau}_1}$ or smaller mixing angles.

\section{Higgs decay rate to two photons}
\label{sec:higgs}
In the MSSM, the effective Higgs coupling to two photons is mediated by the $W$ boson, charged fermions, charged scalar fermions, charged Higgs boson and chargino loops. In this study, among the SUSY particles, we focus on the stau loops, which can give rise to a significant contribution to the two photon decay width \cite{Higgstodiphoton}, together with the dominant $W$ boson and the top quark loops.

Before discussing the stau effects in detail, we briefly explain why we expect a large positive contribution from the staus and why the chargino contribution is  expected to be small even for a lighter chargino mass close to the LEP bound. The two photon decay width of the lighter CP even Higgs boson in the decoupling limit through the charged scalar fermions and charginos loops in addition to the dominant $W$ and the top loops is given by~\cite{Spira}:
\begin{align}
\Gamma(h\to \gamma\gamma)=\frac{G_F\alpha^2m_{h}^3}{128\sqrt{2}\pi^2}
\biggl| A_1(\tau_W) + \frac{4}{3} A_{1/2}(\tau_t)
+C_{\tilde{f}}Q_{\tilde{f}}^2\sum_{i=1}^2 g_{\tilde{f}_i}^{h}A_0(\tau_{\tilde{f}_i}) 
+ \sum_{i=1}^2 g_{\tilde{\chi}_i^-}^{h}A_{1/2}(\tau_{\tilde{\chi}^-_i})
\biggr|^2, \label{width}
\end{align}
where $A_1$, $A_{1/2}$ and $A_0$ denote the spin $1$, $1/2$ and $0$ amplitudes, respectively, and are expressed at the leading order by the following functions:
\begin{subequations}\label{loopamplitude}
\begin{align}
A_1(\tau)&=-
\bigl[
2\tau^2+ 3\tau + 3(2\tau-1)\arcsin^2{\sqrt{\tau}}
\bigr]/\tau^2 
= -7 - \frac{22}{15}\tau + O(\tau^2)\\
A_{1/2}(\tau)&= 2\bigl[
\tau+(\tau-1)\arcsin^2{\sqrt{\tau}}
\bigr]/\tau^2
= \frac{4}{3} + \frac{14}{45}\tau + O(\tau^2) \\
A_0(\tau)&=-
\bigl[
\tau-\arcsin^2{\sqrt{\tau}}
\bigr]/\tau^2
= \frac{1}{3} + \frac{8}{45}\tau + O(\tau^2) .
\end{align}
\end{subequations}
The parameter $\tau_x=m_{h}^2/4m_x^2 (\leq 1)$ is defined by the corresponding masse $m_x$ of the loop particle. The expansion of the loop amplitudes up to $O(\tau)$ tells us that for $m_h\simeq 125$ GeV, the $m_x$ dependence is already small for $m_x \gtrsim 100$ GeV. In eq.~(\ref{width}), $C_{\tilde{f}}$ and $Q_{\tilde{f}}$ denote the color and electromagnetic charges of the sfermion $\tilde{f}$ respectively. The dimensionless coefficients $g$ in front of the loop amplitudes denote the couplings of the Higgs $h$ to the pair of loop particles. Both the spin of a loop particle and the sign of the coefficient $g$, therefore, decide the interference with the dominant $W$ boson loop. 
In principle, how a loop particle affects the Higgs decay rate depends on whether its mass is increased or decreased by the existence of the Higgs field~\cite{LowEnergy}. Let us first examine the chargino contribution. Since the charginos are fermions, they should work constructively with the top quark if their masses are increased by the Higgs field.
The charginos are the mass eigenstates of the winos $\tilde{W}^{1,2}$, each having a soft SUSY breaking mass $M_2$, and the charged Higgsinos $\tilde{H}_d^-$, $\tilde{H}_u^-$, with a supersymmetric Higgsino mass $\mu$. The mixing of these current states is caused by the electroweak symmetry breaking when the Higgs fields acquire vacuum expectation values (VEV). Since the mass of the lighter chargino is always decreased by the mixing, while that of the heavier chargino is always increased, we may expect that the effect of the two charginos tend to cancel each other.  For $\tan{\beta} \gg 1$, the dimensionless coefficients $g_{\tilde{\chi}_i^-}^{h}$ in eq. (\ref{width}), for $|M_2|<|\mu|$,  are approximately given by
\begin{align}
g_{\tilde{\chi}^-_1}^h \propto  -\frac{v}{m_{\tilde{\chi}^-_1}}\frac{m_W|M_2|}{\mu^2-M_2^2},\ \ \ \ 
g_{\tilde{\chi}^-_2}^h \propto  +\frac{v}{m_{\tilde{\chi}^-_2}}\frac{m_W|\mu|}{\mu^2-M_2^2},
\end{align}
while for $|\mu|<|M_2|$,
\begin{align}
g_{\tilde{\chi}^-_1}^h \propto  -\frac{v}{m_{\tilde{\chi}^-_1}}\frac{m_W|\mu|}{M_2^2-\mu^2},\ \ \ \ 
g_{\tilde{\chi}^-_2}^h \propto  +\frac{v}{m_{\tilde{\chi}^-_2}}\frac{m_W|M_2|}{M_2^2-\mu^2}.
\end{align}
These approximate forms of the coefficients show that the relationship $g_{\tilde{\chi}_1}^h \simeq -g_{\tilde{\chi}_2}^h$ can always be satisfied, and since the dependence of the  loop amplitudes on the loop particle mass is small for $m \gtrsim 100$ GeV (eq. (\ref{loopamplitude})), the effects of the two charginos are comparable, and tend to cancel each other. Therefore, the chargino contribution is negligible even when the mass of the lighter chargino is close to the LEP bound.

We now turn to the stau contribution. The mass terms for the stau current eigenstates $\tilde{\tau}_{L, R}$ are given by
\begin{subequations}\label{staumass2}
\begin{align}
m_{\tilde{\tau}_L^{}}^2 &= m_{\tilde{L}}^2 + m_Z^2\cos2\beta\bigl(-1/2+\sin^2\theta_w\bigr) + m_{\tau}^2,\\
m_{\tilde{\tau}_R^{}}^2 &= m_{\tilde{E}}^2 - m_Z^2\cos2\beta \sin^2\theta_w + m_{\tau}^2,
\end{align}
\end{subequations}
where $m_{\tilde{L}}^2$ and $m_{\tilde{E}}^2$ are the soft SUSY breaking masses.  The 2nd and 3rd terms in eqs. (\ref{staumass2}) come from the supersymmetric D- and F- terms, respectively, once the Higgs fields acquire VEV's. In the absence of mixing between $\tilde{\tau}_L$ and $\tilde{\tau}_R$, the mass eigenvalues are given by $m_{\tilde{\tau}_L}^2$ and $m_{\tilde{\tau}_R}^2$. For $\beta > \pi/4$, $m_{\tilde{\tau}_L}^2$ and $m_{\tilde{\tau}_R}^2$ are increased by the 2nd and 3rd terms in eqs. (\ref{staumass2}). From this fact and $A_1(\tau) \times A_0(\tau) < 0$ as in eq. (\ref{loopamplitude}), we can naively expect that both $\tilde{\tau}_L$ and $\tilde{\tau}_R$ contribute to the two photon decay width in the opposite direction to the dominant $W$ boson loop, resulting in a suppression of the decay rate. This changes once we consider the mixing between $\tilde{\tau}_L^{}$ and $\tilde{\tau}_R^{}$, 
\begin{align}
m_{\tilde{\tau}_{LR}^{}}^2 = m_{\tau}\bigl(A_{\tau} - \mu \tan\beta\bigr),\label{mlrstau}
\end{align}
where the 1st and 2nd terms arise, after the electroweak symmetry breaking, from the soft SUSY breaking term and the supersymmetric F-term respectively. If the mass of the lightest stau $m_{\tilde{\tau}_{1}}$ is decreased by $m_{\tilde{\tau}_{LR}}^2$ more than the amount of its increase by the 2nd and 3rd terms in eqs. (\ref{staumass2}), $\tilde{\tau}_{1}$ can work constructively with the dominant $W$ loop, resulting in an enhancement of the decay rate. This enhancement by $\tilde{\tau}_{1}$ depends only on the magnitude of $m_{\tilde{\tau}_{LR}}^2$, and not on the overall sign. However, using a similar argument, the heavier stau $\tilde{\tau}_{2}$ cannot work constructively with the $W$ boson loop, and will thus suppress the decay rate. 

The exact form of the Higgs coupling to the staus,  $g_{\tilde{\tau}_i}^{h}$, can be expressed as \cite{Djouadi}
\begin{subequations}\label{ghstaustau}
\begin{align}
g_{\tilde{\tau}_1}^{h} = \frac{1}{m_{\tilde{\tau}_1}^2}
\biggl[
m_{\tau}^2 + m_Z^2\Bigl(-\frac{1}{2}\cos^2\theta_{\tilde{\tau}} + \sin^2\theta_W \cos2\theta_{\tilde{\tau}} \Bigr)\cos2\beta - \frac{m_{\tau}}{2}\bigl(A_{\tau} - \mu \tan\beta\bigr) \sin2\theta_{\tilde{\tau}}
\biggr],\\
g_{\tilde{\tau}_2}^{h} = \frac{1}{m_{\tilde{\tau}_2}^2}
\biggl[
m_{\tau}^2 + m_Z^2\Bigl(-\frac{1}{2}\sin^2\theta_{\tilde{\tau}} - \sin^2\theta_W \cos2\theta_{\tilde{\tau}} \Bigr)\cos2\beta + \frac{m_{\tau}}{2}\bigl(A_{\tau} - \mu \tan\beta\bigr) \sin2\theta_{\tilde{\tau}}
\biggr],
\end{align}
\end{subequations}
where, $\sin{2\theta_{\tilde{\tau}}}$ is given by
\begin{align}
\sin{2\theta_{\tilde{\tau}}} = \frac{2 m_{\tau} \bigl(A_{\tau} - \mu \tan\beta\bigr)}{m_{\tilde{\tau}_2}^2 - m_{\tilde{\tau}_1}^2}\label{sin2theta_stau}.
\end{align}
The combination  $(A_{\tau} - \mu \tan\beta\bigr) \sin2\theta_{\tilde{\tau}}$ in $g_{\tilde{\tau}_i}^{h}$ of eqs. (\ref{ghstaustau})  is positive definite. Expanding the couplings up to $O(m_{\mathrm{EW}}/m_{\mathrm{SUSY}})$, where $m_{\mathrm{EW}}$ is $m_{\tau}$ or $m_Z$ and $m_{\mathrm{SUSY}}$ is $A_{\tau}$ or $\mu$, and using eq. (\ref{sin2theta_stau}), the stau contribution to the amplitude can be expressed as
\begin{align}
{\cal M} = -\frac{1}{4}\sin^2{2\theta_{\tau}}\bigl( m_{\tilde{\tau}_2}^2 - m_{\tilde{\tau}_1}^2 \bigr)  \Bigl( \frac{1}{m_{\tilde{\tau}_1}^2} A_0(\tau_{\tilde{\tau}_1}) - \frac{1}{m_{\tilde{\tau}_2}^2} A_0(\tau_{\tilde{\tau}_2})
\Bigr) + O\bigl((m_{\mathrm{EW}}/m_{\mathrm{SUSY}})^2\bigr). 
\label{simpleform1}
\end{align}
Recalling that $A_1(\tau)\times A_0(\tau) < 0$ from eq. (\ref{loopamplitude}), the contribution of the lighter stau loop interferes constructively with the dominant $W$ boson loop and can increase the size of the two photon decay width, while the heavier stau loop interferes destructively and can decrease it. However, unlike in the chargino case, the relationship $g_{\tilde{\tau}_1}^{h} \sim - g_{\tilde{\tau}_2}^{h}$ is not necessary satisfied. Both the large mixing angle $|\theta_{\tilde{\tau}}| \sim 45^{\circ}$ and large mass splitting $m_{\tilde{\tau}_2}-m_{\tilde{\tau}_1}$ imply $g_{\tilde{\tau}_1}^{h} \gg - g_{\tilde{\tau}_2}^{h}$, thus resulting in an enhancement of $\Gamma(h\to \gamma\gamma)$ by the lightest stau $\tilde{\tau}_1$.

In Figure~\ref{plotdm1} we show contours of the ratio $\Gamma(h \rightarrow \gamma \gamma)/\Gamma (H_{SM}^{} \rightarrow \gamma \gamma)$ in the $\mla-\th$ plane for different values of $\Delta m_{\tilde{\tau}}$. For our numerical analysis, we have taken the model parameters as $m_t=173.5$, $m_{\tau}=1.77682$, $m_W=80.385$, $m_Z=91.1876$, $m_h=125.5$ GeV, $\sin^2{\theta_w}=0.23116$ and $\tan{\beta}=10$. This ratio can be compared with the measured signal strengths at the LHC, assuming that the Higgs production cross-section and total decay widths are not significantly modified. The present ATLAS measurement of the Higgs signal strength in the di-photon channel is $1.6 \pm 0.3$~\cite{ATLAS-gaga}, while the CMS measurement is $0.78 \pm 0.27$  from their MVA analysis and $1.11 \pm 0.31$ from their cut-based analysis~\cite{CMS-gaga}. Since the SM value of $1$ is consistent at the $2\sigma$ level with both the current measurements, no conclusion can be drawn at the moment from the data. However, since the ATLAS data continues to show an enhancement in the central value, future improvement of this measurement may provide the first hint of new physics in Higgs properties.

From Figure~\ref{plotdm1}, we can see the correlation between this decay rate and the neutralino annihilation rate to tau pairs. Two important differences between these two observables are that even though the di-photon branching fraction also increases with $\Delta m_{\tilde{\tau}}$, but unlike in the case of $\langle \sigma v \rangle(\neu \neu \rightarrow \tau^+ \tau^-)$, it does not reach an asymptotic value for some $\Delta m_{\tilde{\tau}}$. In addition, the diphoton branching fraction is symmetric about $\th=0^{\circ}$, while $\langle \sigma v \rangle$ is not. We emphasize that the region of the $\tilde{\tau}$ parameter space favored by the DM annihilation rate $\langle \sigma v\rangle\sim \langle\sigma v\rangle_0$ overlaps with the region which enhances the $h\to \gamma\gamma$ rate.

\section{Relic density of a 10 GeV $\neu$ DM}
\label{sec:relic}
The current estimate for DM relic density, including recent data from Planck is $\Omega_{\rm DM} h^2=0.1199 \pm 0.0027$ ($68\%$ C.L.)~\cite{WMAP,Planck}, which amounts to a thermally averaged annihilation cross-section at the time of freeze-out $\langle \sigma v\rangle_{\rm F.O.} \simeq 2.2 \times 10^{-26}{~\rm cm^3}$/s~\cite{Steigman}. For a light $\mathcal{O}(10 {~\rm GeV})$ bino-like $\neu$ DM, $\langle \sigma v\rangle_{\rm F.O.}$ is found to be lower than the above value in large regions of MSSM parameter space, leading to constraints on otherwise allowed parameter regions, and in particular lower bounds on the $\neu$ mass~\cite{relic-bound}. Recently, it was argued in Ref.~\cite{Boehm-recent} that a 10 GeV $\neu$ can satisfy the relic density constraint if all the slepton masses are close to the LEP bound. However, the LHC bounds on the first two generation sleptons have become stronger since then~\cite{LHCslepton}, and when combined with the LEP results~\cite{stau1,stau2,stau4} which are sensitive in a complementary region, it is difficult to have selectrons or smuons lighter than 230 GeV (for details, see section~\ref{slepton}). 

Since the coupling of a  bino-like $\neu$ to sleptons is proportional to their hypercharge, the contribution of $\tilde{\ell}_R$ ($\ell=e,\mu$) to $\langle \sigma v\rangle_{\rm F.O.}$ is $16$ times larger than that of a $\tilde{\ell}_L$ of the same mass. Combining this with the fact that the ATLAS bound on $\tilde{\ell}_L$ is stronger ($\tilde{\ell}_L>300$ GeV), we conclude that for achieving a higher value of $\langle \sigma v\rangle_{\rm F.O.}$, it is better to interpret the ATLAS bound as $\tilde{\ell}_R>230$ GeV with $\tilde{\ell}_L$ decoupled, rather than taking the common bound of 320 GeV on all first two generation sleptons. We have also verified this numerically. Even then, the contribution of $\tilde{\ell}_R$ to $\langle \sigma v\rangle_{\rm F.O.}$ is rather low, $\mathcal{O}(10^{-28}{~\rm cm^3/s})$. Therefore, we conclude that in view of the combined LEP and recent ATLAS limits, the first two generation sleptons do not contribute significantly to $\langle \sigma v\rangle_{\rm F.O.}$ for a 10 GeV $\neu$.

In order to evaluate the stau contribution, let us first fix $\mla,\mlb$ and $\th$ such that $\langle \sigma v \rangle(\neu \neu \rightarrow \tau^+ \tau^-)=\langle \sigma v \rangle_0$ in the $v\rightarrow 0$ limit, where $\langle \sigma v \rangle_0=2 \times 10^{-27}{~\rm cm^3/s}$ (eq.~\ref{eqn-value}) is the best-fit value obtained by fitting the Fermi Bubble data. In the maximal mixing case, $\th=45^\circ$, a possible solution satisfying the LEP bounds is $\mla = 90$ GeV and $\mlb=131.5$ GeV, with $m_{\tilde{\nu}_\tau}=80.1$ GeV. With this choice, for a pure bino-like $\neu$, we find (using {\tt micrOMEGAs-3.2}) that $\Omega_{\rm DM} h^2 = 0.808$, with $93\%$ contribution to $\langle \sigma v \rangle_{\rm F.O.}$ coming from the $\tau^+\tau^-$ channel, and the rest from the $\nu_\tau \bar{\nu}_\tau$, $e^+e^-$ and $\mu^+\mu^-$ channels. Obeying the current bounds on slepton and sneutrino masses discussed in section~\ref{slepton}, it does not seem to be possible to achieve the required value of $2.2 \times 10^{-26}{~\rm cm^3/s}$ for $\langle \sigma v \rangle_{\rm F.O.}$. Since the LHC mass bounds on the first two generation squarks is around 1 TeV~\cite{ATLAS-squark}, and that on sbottoms  is around 640 GeV~\cite{ATLAS-sbottom},  the contribution from the $b\bar{b}$ and $c\bar{c}$ final states for the $\neu$ pair-annihilation are also not significant, while the loop-induced contribution to the $gg$ final state is suppressed by a factor of $\alpha_s^2$ and can be neglected in relic abundance calculations compared to the fermionic final states~\cite{Jungman}. 
The DM density in our scenario is thus higher than the measured value, and therefore, if we use a standard thermal history of the universe, DM will be over-abundant.

However, there are well-motivated non-standard possibilities for the thermal history of the universe. In such scenarios, late-time decays of heavy fields can produce relativistic particles and entropy, thereby diluting the relic density of DM produced at the time of freeze-out. A particularly motivated example of such long-lived massive particles is moduli. The generic requirements in such a scenario are that: 1) the decay of the heavy particle should reheat the universe to a temperature ($T_{\rm R.H.}$) which is below the DM freeze-out temperature ($T_{\rm F.O.}$); 2) $T_{\rm R.H.}$ should be higher than the temperature scale in which big-bang nucleosynthesis (BBN) takes place; and 3) In a SUSY theory, the decay branching fraction of the heavy particle to gauginos (especially, in our case, $\neu$), should be very small, so as not to produce more DM from the decay. There are generic models which satisfy all these conditions, and for a recent study  focussing on the solution to the DM over-abundance problem, see Ref.~\cite{Hooper-moduli}. For $T_{\rm R.H.}<T_{\rm F.O.}$, the relic density of DM is diluted by the factor $(T_{\rm R.H.}/T_{\rm F.O.})^3$. Therefore for the parameter choice described above, predicting $\Omega_{\rm DM} h^2 = 0.808$, to reduce the relic density to within $2\sigma$ of the Planck result, we need to have $T_{\rm R.H.}=0.27$ GeV ($T_{\rm F.O.} \simeq m_{\neu}/20 = 0.5$ GeV). As shown in Ref.~\cite{Hooper-moduli}, this value of $T_{\rm R.H.}$ is sufficiently far away from the BBN epoch and requires a moduli field of mass around 1 PeV. The branching fraction of such a moduli field to SUSY particles is also negligibly small, $\mathcal{O}(10^{-8})$, thereby satisfying all the requirements. Therefore, in the MSSM parameter region of our interest, the DM over-abundance problem can be evaded by appealing to a non-standard thermal history of the universe.

\section{The muon $g-2$ with a 10 GeV neutralino}
\label{sec:muon}
The difference between the SM prediction and the experimental value of the muon $g-2$ is given by \cite{g-2_1}
\begin{align}
a_{\mu}^{EXP}-a_{\mu}^{SM} = (26.1\pm8.0) \times 10^{-10} \label{g-2data}
\end{align}
which differs from zero at $3.3\sigma$, where $a_{\mu}=(g_{\mu}-2)/2$. We take this discrepancy as a hint of new physics, and explore to what extent it can be accounted for by the supersymmetric particles. In the MSSM, $a_{\mu}$ receives contributions at one loop level from the chargino $(\tilde{\chi}^{\pm})-$muon sneutrino $(\tilde{\nu}_{\mu})$ loop and the neutralino $(\tilde{\chi}^0)-$smuon $(\tilde{\mu})$ loop.

In this section, we estimate the MSSM contribution to the muon $g-2$ in our scenario with a $10$ GeV bino-like neutralino, taking into account all the experimental constraints discussed in Section 2. Our main findings can be summarized as follows. Even though the bino-like neutralino is significantly light, the bino contribution is unlikely to account for the difference in eq. (\ref{g-2data}), unless $(A_{\mu}-\mu\tan{\beta}) \sim 1770$ TeV for $m_{\tilde{\mu}_1}=350$ GeV and $\theta_{\tilde{\mu}}=45^{\circ}$. After taking into account the recent ATLAS bounds on chargino and smuon masses, we find that the contribution from the chargino-muon sneutrino loop is sufficient to explain the data, even if the contributions from the other diagrams are small. However, in order to accommodate the $g-2$ discrepancy within $1\sigma$ a somewhat larger value of $\tan{\beta}$ (greater than 20) is necessary.

Following the paper by G. C. Cho et al. \cite{GCcho}, the leading terms of the MSSM contribution to the muon $g-2$ in the $m_{\mathrm{EW}}/m_{\mathrm{SUSY}}$ expansion, where $m_{\mathrm{EW}}$ is $m_{\mu}$, $m_W$ or $m_Z$ and $m_{\mathrm{SUSY}}$ is $m_{\tilde{\mu}}$, $m_{\tilde{\nu}}$, $M_1$, $M_2$ or $\mu$, are given by five diagrams written in terms of weak eigenstates, see Figure 1 and eq. (2.6) in ref. \cite{GCcho}. In our scenario, where the bino and Higgsino must be decoupled from each other, the dominant terms are given by
\begin{subequations}\label{g-2approximation}
\begin{align}
\Delta a_{\mu}(\tilde{W}-\tilde{H}, \tilde{\nu}_{\mu})&=\frac{g^2m_{\mu}^2}{8\pi^2}\frac{M_2\mu\tan\beta}{m_{\tilde{\nu}_{\mu}}^4}
F_a(M_2^2/m_{\tilde{\nu}_{\mu}}^2,\mu^2/m_{\tilde{\nu}_{\mu}}^2),\label{1st}\\
\Delta a_{\mu}(\tilde{W}-\tilde{H}, \tilde{\mu}_L)&=-\frac{g^2m_{\mu}^2}{16\pi^2}\frac{M_2\mu\tan\beta}{m_{\tilde{\mu}_L}^4}
F_b(M_2^2/m_{\tilde{\mu}_L}^2,\mu^2/m_{\tilde{\mu}_L}^2),\label{d}\\
\Delta a_{\mu}(\tilde{B}, \tilde{\mu}_L-\tilde{\mu}_R)&=-\frac{g_Y^2m_{\mu}^2}{16\pi^2}
\frac{A_{\mu}-\mu\tan{\beta}}{M_1^3}
F_b(m_{\tilde{\mu}_1}^2/M_1^2, m_{\tilde{\mu}_2}^2/M_1^2)\nonumber \\
&= -\frac{g_Y^2m_{\mu}}{32\pi^2}
\frac{(m_{\tilde{\mu}_2}^2- m_{\tilde{\mu}_1}^2)
\sin{2\theta_{\tilde{\mu}}}} {M_1^3}
F_b(m_{\tilde{\mu}_1}^2/M_1^2, m_{\tilde{\mu}_2}^2/M_1^2),\label{2nd}
\end{align}
\end{subequations}
where 
\begin{subequations}
\begin{align}
&F_a(x,y)=-\frac{G_3(x) - G_3(y)}{x-y},\\
&F_b(x,y)=-\frac{G_4(x) - G_4(y)}{x-y},\\
&G_3(x)=\frac{1}{2(x-1)^3}[(x-1)(x-3)+2\ln{x}],\\
&G_4(x)=\frac{1}{2(x-1)^3}[(x-1)(x+1)-2x\ln{x}].
\end{align}
\end{subequations}
Note that $F_a(x,y)$ and $F_b(x,y)$ are defined to be positive for all positive $x$ and $y$, and $F_a(x,y)$ is always larger than $F_b(x,y)$ for the same arguments \cite{GCcho}.  Eq. (\ref{1st}) expresses the contribution from the wino-Higgsino mixed charginos and muon-sneutrino loop, while  eq. (\ref{d}) from the wino - Higgsino mixed neutralinos and left-handed smuon loop. These two contributions can be efficient when the wino and Higgsino are maximally mixed, namely $|M_2/\mu|\sim1$. If we neglect the electroweak terms in the left-handed smuon and the muon sneutrino masses, both of them are equal to $m_{\tilde{L}}^2$ and the arguments of $F_{a,b}(x,y)$ in eqs. (\ref{1st}, \ref{d}) are the same. Considering a factor of $2$ and $F_a(x,y) > F_b(x,y)$, we can conclude that the positive discrepancy between the data and the SM prediction favors $M_2\mu > 0$. We emphasize that the right handed smuon is not relevant to these processes. Eq. (\ref{2nd}) expresses the contribution from the bino  and left-right mixed smuon loop, which is positive for $(\mu\tan{\beta}-A_{\mu})M_1 > 0$, which implies $M_1\mu > 0$.

\begin{figure}[t]
\centering
\includegraphics[scale=0.81]{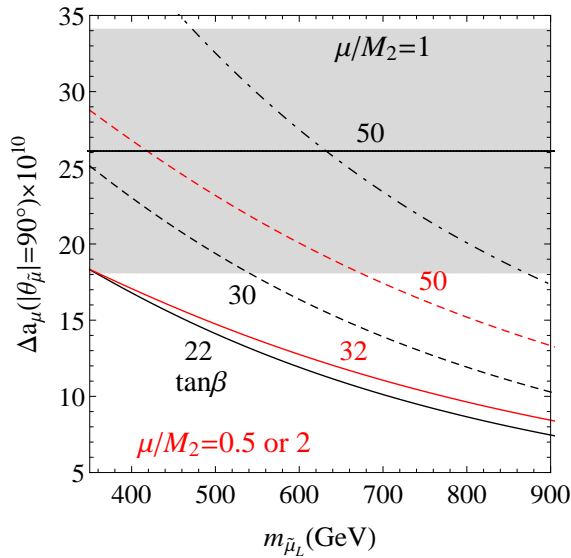}
\caption{$\Delta a_{\mu}$ as a function of the left handed smuon mass $m_{\tilde{\mu}_L}$ when the smuon mixing is fixed to zero, $|\theta_{\tilde{\mu}}|=90^{\circ}$.
The mass of the lighter chargino is fixed at $m_{\tilde{\chi}^-_1}=350$ GeV. See text in Section \ref{sec:muon} for a detailed explanation of the plot.}
\label{plot8}
\end{figure}

We first consider a scenario in which the smuon mixing angle $|\theta_{\tilde{\mu}}|$ is set to $90^{\circ}$. This is achieved when,  in eq. (\ref{massmatrix}),
\begin{align}
m^2_{\tilde{\mu}_L^{}}- m^2_{\tilde{\mu}_R^{}} \gg |2m^2_{\tilde{\mu}_{LR}^{}}|.
\end{align}
Since $m^2_{\tilde{\mu}_{LR}^{}}$ is proportional to the muon mass $m_{\mu}$,  this relationship is satisfied except when $m^2_{\tilde{\mu}_R^{}} = m^2_{\tilde{\mu}_L^{}}$ or $|A_{\mu}-\mu\tan{\beta}|$ is very large. For $|\theta_{\tilde{\mu}}|=90^{\circ}$, according to eqs. (\ref{g-2approximation}),  the chargino and muon sneutrino loop dominantly contributes to $\Delta a_{\mu}$, whereas the contribution involving the bino loop is suppressed. In Figure \ref{plot8}, we show the MSSM contribution to $\Delta a_{\mu}$ as a function of the left handed smuon mass $m_{\tilde{\mu}_L}$ with $|\theta_{\tilde{\mu}}|=90^{\circ}$. The mass of the lighter chargino is fixed at $m_{\tilde{\chi}^-_1}=350$ GeV, which is  just above the LHC bound quoted in Section \ref{slepton}. Along the horizontal axis we show the mass of $\tilde{\mu}_L$, since the contribution of the right-handed smuon is very small in the zero mixing case, which is clearly seen in eqs. (\ref{g-2approximation}). The smuon mass $m_{\tilde{\mu}_L}$ is varied from $m_{\tilde{\mu}_L}=m_{\tilde{\chi}^-_1}=350$ GeV, which prevents the lighter chargino from decaying via the smuon, see Section \ref{slepton}. For our numerical analysis, we have fixed $m_{\tilde{\mu}_{R}}=350$ GeV for definiteness, and changing this value does not modify our results. The discrepancy between the data and the SM prediction quoted in eq. (\ref{g-2data}) is indicated by the horizontal solid black line, while the shaded region corresponds to the $1\sigma$ uncertainty. We have fixed the value of $\tan\beta$ along the curves. For the black curves, $M_2/\mu = 1$ is satisfied, where the wino and Higgsino are maximally mixed and the largest contribution is predicted. On the other hand, for the red curves, $M_2/\mu = 0.5$ or $2$ is satisfied. $M_2/\mu = 0.5$ represents the case when the lighter chargino is wino-like, while $M_2/\mu = 2$ corresponds to a  Higgsino-like lighter chargino. The result is symmetric between these two cases, as also implied by eqs. (\ref{1st}) and (\ref{d}). The signs of $M_1$ and $M_2$ are chosen such that $M_2\mu>0$ and $M_1\mu>0$. Figure~\ref{plot8} shows that the contribution from only the chargino and muon-sneutrino loop is sufficient to explain the discrepancy in the muon $g-2$ within $1\sigma$ only when $\tan{\beta}$ is large, at least $22$ for $M_2/\mu = 1$ and $32$ for $M_2/\mu = 0.5$ or $2$. For the analysis in this plot, we have used the full one loop result given in ref. \cite{g-2SUSY}.

Next, we consider a scenario in which the smuon mixing is maximal $\theta_{\tilde{\mu}} = 45^{\circ}$ and estimate the contribution from the bino and left-right mixed smuon loop given by eq. (\ref{2nd}). The maximal mixing is obtained for $m^2_{\tilde{\mu}_R^{}} = m^2_{\tilde{\mu}_L^{}}$ with non-zero values of $(A_{\mu}-\mu\tan{\beta})$. We can see from  eq. (\ref{2nd}) that the contribution involving the bino loop  increases approximately linearly with $(A_{\mu}-\mu\tan{\beta})$. If we assume universal soft trilinear terms for all sleptons, then $A_{\mu}-\mu\tan{\beta} \simeq A_{\tau}-\mu\tan{\beta}$. From Figure \ref{plotdm1}, in the region of the $\tilde{\tau}$ parameter space favoured by the DM annihilation rate, $(A_{\tau}-\mu\tan{\beta})$ takes values between $+3$ TeV and $+47$ TeV. For $m_{\tilde{\mu}_1^{}}=350$ GeV and $\theta_{\tilde{\mu}} = +45^{\circ}$, the value of $A_{\mu}-\mu\tan{\beta}=3$ TeV gives $\Delta a_{\mu}(\tilde{B}, \tilde{\mu}_L-\tilde{\mu}_R)\simeq +0.2$, and the value of $A_{\mu}-\mu\tan{\beta}=47$ TeV gives $\Delta a_{\mu}(\tilde{B}, \tilde{\mu}_L-\tilde{\mu}_R)\simeq +2.6$, both in units of $10^{-10}$. These are evaluated using eq. (\ref{2nd}), where we have set $M_1=-10$ GeV such  that $\Delta a_{\mu}(\tilde{B}, \tilde{\mu}_L-\tilde{\mu}_R)$ gives a positive contribution. Thus we find that the contribution from the bino and left-right mixed smuon loop cannot contribute significantly to the difference (eq. ~\ref{g-2data}) even for a very light bino ($10$ GeV), if the trilinear term $A_\mu$ is of the same order as $A_\tau$ which gives the desired bino annihilation rate.

\section{Summary}
\label{sec:summary}
To summarize our study, we consider the possibility that a component of the low-latitude emission in the Fermi Bubble gamma ray data can originate from a 10 GeV DM particle annihilating to $\tau^+ \tau^-$ with $\langle \sigma v \rangle=2 \times 10^{-27} {~\rm cm^3/s}$. We explore $\neu$ pair-annihilation to $\tau^+\tau^-$ via stau exchange diagrams, and determine the region in the ($\mla,\mlb,\th$) parameter space which can accommodate the above data. It is found that the stau mixing plays a major role in enhancing $\langle \sigma v \rangle$, and for a given $\mla$ and $\th$, $\langle \sigma v \rangle$ increases with $\Delta m_{\tilde{\tau}} = \mlb - \mla$,  reaching an asymptotic value at around $\Delta m_{\tilde{\tau}}=200$ GeV. To determine the constraints in the stau sector, we revisit the LEP direct search bound on $\mla$ as a function of $\th$. We also use the electroweak precision constraints to find that upto $\Delta m_{\tilde{\tau}}=300$ GeV with a light $\tla$ and significant mixing $\th$, the contribution of the stau loops to the oblique parameters are consistent with the data. We also verified that the necessary stau mass splitting and mixing angle are compatible with the requirements of vacuum meta-stability.
After taking into account the constraints on the Higgsino mass parameter $\mu$, and hence on the Higgsino admixture to a bino-like $\neu$, coming from the upper bounds on the $Z$ and Higgs boson invisible widths  as well as the bound on chargino mass from LHC in a scenario with a light stau, we go on to explore the implications of this small Higgsino fraction to the annihilation rate. 

One of the main results of our paper is the correlation of $\langle \sigma v \rangle (\neu \neu \rightarrow \tau^+ \tau^-)$ with $\Gamma(h\rightarrow \gamma \gamma)$, since both these observables depend crucially on $\mla,\th$ and $\Delta m_{\tilde{\tau}}$. We find that both are enhanced in presence of a light $\tla$ and large mixing $\th$, and there are regions in the $\mla-\th$ parameter space which can accommodate both the required annihilation rate as well as the possibility of an enhanced Higgs decay rate to two photons. In particular, for $\tla$ masses obeying the most stringent lower bound from LEP (OPAL), in the parameter region consistent with the best-fit value of the annihilation rate, $\Gamma(h\rightarrow \gamma \gamma)$ is always found to be enhanced. The dependence on the sign of $\th$ is however found to be different.  For the same $\mla$ and $\Delta m_{\tilde{\tau}}$, $\langle \sigma v \rangle$ is larger for $\th>0$, but $\Gamma(h\rightarrow \gamma \gamma)$ is symmetric about $\th=0$. 

Since the allowed Higgsino fraction of a 10 GeV $\neu$ is rather small, and the recent ATLAS SUSY search results put stringent bounds on the first two generation sleptons ($m_{\tilde{e}_R},m_{\tilde{\mu}_R}>230$ GeV), we find that it is difficult to achieve a sufficiently large annihilation rate $\langle \sigma v\rangle_{\rm F.O.}$, required to reproduce the observed value of the DM relic density. To avoid the problem of DM over-abundance, a non-standard thermal history of the universe may be required, in which the late time decay of a heavy field like moduli reheats the universe to a temperature below the DM freeze-out temperature. Taking an example MSSM mass spectrum in our scenario, we find that the required value is of the order of $T_{\rm R.H.}=0.27$ GeV, which is sufficiently above the epoch of BBN.  A decay of a moduli field of mass around 1 PeV can give rise to such a $T_{\rm R.H.}$, and the branching fraction of such a moduli field to SUSY particles is also negligibly small, $\mathcal{O}(10^{-8})$. Since in many well-motivated theoretical scenarios the thermal history of the universe can be modified, the MSSM parameter space explored in this study can be made consistent with the relic density requirement. 

Finally, we estimate the MSSM contribution to the muon $g-2$ taking into account all the experimental constraints on the MSSM spectrum, in particular, the recent ATLAS bounds on the smuon and chargino masses.  We find that even though the bino-like neutralino is significantly light, the bino contribution is unlikely to account for the discrepancy. However, it is possible to explain the anomaly within $1\sigma$ by the chargino and muon-sneutrino loop alone with a somewhat larger value of $\tan \beta$ ($>20$).

Therefore, it will be interesting to look for signals of a light stau with a mass less than about 130 GeV in the future LHC data, in final states with boosted tau leptons and missing energy. If in addition, with the accumulation of large statistics, the LSP and NLSP masses could be estimated, that would lead to a direct test of the scenario considered in our study. Indirect hints might also be forthcoming in the future LHC measurement of the $h \rightarrow \gamma \gamma$ signal strength with a much better accuracy. It is also very important to see if a similar hint is found in the future Fermi-LAT gamma ray data from dwarf spheroidal galaxies, which are much cleaner probes for studying signals of annihilating DM.

\section*{Acknowledgments}
We wish to thank Rafael Lang for stimulating discussions on the subject. JN is grateful to Daisuke Nomura for his help in calculations of the oblique corrections and the muon $g-2$. This work is supported by the World Premier International Research Center Initiative (WPI Initiative), MEXT, Japan, for SM, and by JSPS KAKENHI Grant Number 20340064, 25$\cdot$4461, for KH and JN. 

\appendix
 \renewcommand{\theequation}{A-\arabic{equation}}
  \setcounter{equation}{0}  

\section*{Appendix}

\section{Neutralino mass and mixing}\label{app1}
The neutralino mass term in the current basis $X_L = (\tilde{B}_L, \tilde{W}_L^{3}, \tilde{H}_{d L}^{0}, \tilde{H}_{u L}^{0})^T$ and $X_R = (\tilde{B}_R, \tilde{W}_R^{3}, \tilde{H}_{d R}^{0}, \tilde{H}_{u R}^{0})^T$ is given by
\begin{equation}
- {\cal L} = \frac{1}{2} X_R^{\dagger} M X_L^{} + h.c.,\\
\end{equation}
where, the mass matrix is 
\begin{align}
M =
\begin{pmatrix}
M_1 & 0 & -m_Zs_wc_{\beta} & +m_Zs_ws_{\beta} \\
0 & M_2 & +m_Zc_wc_{\beta} & -m_Zc_ws_{\beta} \\
-m_Zs_wc_{\beta} & +m_Zc_wc_{\beta} & 0 & -\mu \\
+m_Zs_ws_{\beta} & -m_Zc_ws_{\beta} & -\mu & 0
\end{pmatrix}
\end{align}
with abbreviations $s_w=\sin{\theta_w}$, $c_w=\cos{\theta_w}$, $s_{\beta}=\sin{\beta}$ and $c_{\beta}=\cos{\beta}$.  $M_1$, $M_2$ and $\mu$ denote the masses of bino, winos and Higgsinos, respectively. If $M_i$ and $\mu$ are real, the neutralino mass matrix $M$ is real and symmetric, which can be diagonalized by a real unitary (orthogonal) matrix $U$ as
\begin{align}
U^TMU = \mathrm{diag}(m_1, m_2, m_3, m_4),\ \ 0 \leq |m_1| < |m_2| < |m_3| < |m_4|, 
\end{align}
where $m_i$ are real but not necessarily positive. In general the eigenvalues can be expressed as
\begin{align}
m_i = |m_i|e^{i\xi_i}.
\end{align}
We, therefore, introduce a diagonal phase matrix 
\begin{align}
P = \mathrm{diag}(e^{i\frac{\xi_1}{2}}, e^{i\frac{\xi_2}{2}}, e^{i\frac{\xi_3}{2}}, e^{i\frac{\xi_4}{2}})
\end{align}
which makes the neutralino masses positive, and define
\begin{align}
U_L^N = U P^*,\ \ U_R^N = U^*P,
\end{align}
so that 
\begin{subequations}
\begin{align}
U_R^{N \dagger}M U_L^N& = \mathrm{diag}(|m_1|, |m_2|, |m_3|, |m_4|)\\
& = \mathrm{diag}(m_{\tilde{\chi}_1^0}, m_{\tilde{\chi}_2^0}, m_{\tilde{\chi}_3^0}, m_{\tilde{\chi}_4^0}),
\end{align}
\end{subequations}
where $m_{\tilde{\chi}_i^0}$ are now positive and $0 \leq m_{\tilde{\chi}_1^0} < m_{\tilde{\chi}_2^0} < m_{\tilde{\chi}_3^0} < m_{\tilde{\chi}_4^0}$. 
The mass eigenstates $\tilde{\chi}_{i}^0$ are obtained from
\begin{subequations}
\begin{align}
X_{L i}& = (U_L^N)_{ij} \tilde{\chi}_{j L}^0,\\
X_{R i}& = (U_R^N)_{ij} \tilde{\chi}_{j R}^0.
\end{align}
\end{subequations}
Consequently, the mass eigenstates are expressed as
\begin{subequations}
\label{mass-eigen}
\begin{align}
\tilde{\chi}_{i L}^0 &= (U_L^{N \dagger})_{ij}X_{L j} = P_{ii} (U^{\dagger})_{ij} X_{L j} \equiv N_{ij} X_{L j}\\
\tilde{\chi}_{i R}^0 &= (U_R^{N \dagger})_{ij}X_{R j} = (P^*)_{ii} (U^{T})_{ij} X_{R j} = (N^*)_{ij} X_{R j}.
\end{align}
\end{subequations}
It should be noted that in our phase convention, the Majorana conditions for the current states,
\begin{align}
X_{R i} = - i\sigma^2 (X_{L i})^*
\end{align}
are preserved for the mass eigenstates (neutralinos),
\begin{align}
\tilde{\chi}_{i R}^0 = - i\sigma^2 (\tilde{\chi}_{i L}^0)^*.
\end{align}

\section{Slepton mass and mixing}\label{sleptonmixing}
In this appendix, we define our notation for slepton masses and mixing, which play a major role in our study. The scalar supersymmetric partners $\tilde{l}_{L,R}$ of the left and right handed leptons mix with each other and give rise to the mass eigenstates $\tilde{l}_{1,2}$. The current eigenstates $\tilde{l}_{L,R}$ are related to the mass eigenstates $\tilde{l}_{1,2}$ by a real unitary matrix as
\begin{align}
\begin{pmatrix}
\tilde{l}_L\\
\tilde{l}_R
\end{pmatrix}
=
\begin{pmatrix}
\cos\theta_l & \sin\theta_l\\
-\sin\theta_l & \cos\theta_l
\end{pmatrix}
\begin{pmatrix}
\tilde{l}_1\\
\tilde{l}_2
\end{pmatrix}.\label{mixing}
\end{align}
The slepton mass squared matrix  in the current state basis is given by
\begin{align}
-{\cal L} =
\begin{pmatrix}
\tilde{l}_L^* & \tilde{l}_R^*
\end{pmatrix}
\begin{pmatrix}
m_{\tilde{l}_L}^2 & m_{\tilde{l}_{LR}}^2\\
m_{\tilde{l}_{LR}}^{2} & m_{\tilde{l}_R}^2
\end{pmatrix}
\begin{pmatrix}
\tilde{l}_L\\
\tilde{l}_R
\end{pmatrix},\label{massmatrix}
\end{align}
whose matrix elements are 
\begin{subequations}\label{m}
\begin{align}
m_{\tilde{l}_L}^2 &= m_{\tilde{L}}^2 + m_Z^2\cos2\beta\bigl(-1/2+\sin^2\theta_w\bigr) + m_l^2, \label{mLL}\\
m_{\tilde{l}_R}^2 &= m_{\tilde{E}}^2 - m_Z^2\cos2\beta\sin^2\theta_w + m_l^2, \label{mRR}\\
m_{\tilde{l}_{LR}}^2 &= m_l\bigl(A_l - \mu \tan\beta\bigr).\label{mLR}
\end{align}
\end{subequations}
By diagonalizing the mass squared matrix, we determine the mass eigenvalues:
\begin{align}
m^2_{\tilde{l}_1, \tilde{l}_2}= \frac{m_{\tilde{l}_L}^2 + m_{\tilde{l}_R}^2}{2}\mp \frac{\sqrt{(m_{\tilde{l}_L}^2-m_{\tilde{l}_R}^2)^2+(2m_{\tilde{l}_{LR}}^2)^2}}{2}. \label{mass_eigenvalue}
\end{align}
The squared mass of the corresponding sneutrino is given by
\begin{align}
m_{\tilde{\nu}}^2 &= m_{\tilde{L}}^2 + \frac{1}{2}m_Z^2\cos2\beta.
\end{align}
The mixing angle $\theta_{\tilde{l}}$ takes the form
\begin{subequations}\label{theta}
\begin{align}
\sin2\theta_{\tilde{l}}& = \frac{2m_{\tilde{l}_{LR}}^2}{\sqrt{(m_{\tilde{l}_L}^2-m_{\tilde{l}_R}^2)^2+(2m_{\tilde{l}_{LR}}^2)^2}}=\frac{2m_{\tilde{l}_{LR}}^2}{m^2_{\tilde{l}_2} - m^2_{\tilde{l}_1}}, \label{sin2theta}\\
\cos2\theta_{\tilde{l}}& = \frac{-(m_{\tilde{l}_L}^2-m_{\tilde{l}_R}^2)}{\sqrt{(m_{\tilde{l}_L}^2-m_{\tilde{l}_R}^2)^2+(2m_{\tilde{l}_{LR}}^2)^2}} = \frac{-(m_{\tilde{l}_L}^2-m_{\tilde{l}_R}^2)}{m^2_{\tilde{l}_2} - m^2_{\tilde{l}_1}},
\end{align}
\end{subequations}
and, in addition,
\begin{subequations}\label{theta2}
\begin{align}
\sin\theta_{\tilde{l}}& = \frac{m_{\tilde{l}_{LR}}^2}{\sqrt{(m_{\tilde{l}_2}^2-m_{\tilde{l}_L}^2)^2+(m_{\tilde{l}_{LR}}^2)^2}},\\
\cos\theta_{\tilde{l}}& = \frac{m_{\tilde{l}_2}^2-m_{\tilde{l}_L}^2}{\sqrt{(m_{\tilde{l}_2}^2-m_{\tilde{l}_L}^2)^2+(m_{\tilde{l}_{LR}}^2)^2}}
\end{align}
\end{subequations}
are given once the mixing angle $\theta_{\tilde{l}}$ is chosen in the region $-\pi/2 < \theta_{\tilde{l}} < \pi/2$. Note that $m_{\tilde{l}_2}^2-m_{\tilde{l}_L}^2>0$ is always satisfied. The region of $\theta_{\tilde{l}}$ can further be separated by using equations(\ref{theta}, \ref{theta2}) as follows:
\begin{subequations}
\label{octant}
\begin{align}
\mathrm{When}\ & m_{\tilde{l}_R}^2 - m_{\tilde{l}_L}^2 > 0\ \mathrm{and}\ m_{\tilde{l}_{LR}}^2 < 0,\ \mathrm{then}\ -\pi/4 < \theta_{\tilde{l}} < 0\label{theta21}\\
\mathrm{When}\ & m_{\tilde{l}_R}^2 - m_{\tilde{l}_L}^2 < 0\ \mathrm{and}\ m_{\tilde{l}_{LR}}^2 < 0,\ \mathrm{then}\ -\pi/2 < \theta_{\tilde{l}} < - \pi/4\label{theta23}\\
\mathrm{When}\ & m_{\tilde{l}_R}^2 - m_{\tilde{l}_L}^2 > 0\ \mathrm{and}\ m_{\tilde{l}_{LR}}^2 > 0,\ \mathrm{then}\  0 < \theta_{\tilde{l}} < \pi/4\label{theta22}\\
\mathrm{When}\ & m_{\tilde{l}_R}^2 - m_{\tilde{l}_L}^2 < 0\ \mathrm{and}\ m_{\tilde{l}_{LR}}^2 > 0,\ \mathrm{then}\ \pi/4 < \theta_{\tilde{l}} < \pi/2.\label{theta24}
\end{align}
\end{subequations}


\end{document}